\documentclass[twocolumn,preprintnumbers,amsmath,amssymb,superscriptaddress,subeqn,prl]{revtex4-1}

\usepackage[table,xcdraw]{xcolor}
\usepackage{amsmath}
\usepackage{amsthm}
\usepackage{float}
\usepackage{xcolor}
\usepackage{dsfont}

\usepackage{multirow}

\usepackage{tabularx} 
\newcolumntype{Y}{>{\centering\arraybackslash}X}
\newcolumntype{B}{>{\centering\columnwidth=2\columnwidth}Y}
\newcolumntype{S}{>{\centering\columnwidth=.75\columnwidth}Y}

\usepackage{graphicx} 
\usepackage{times}
\usepackage{footmisc}
\usepackage{soul}
\usepackage{color} 
\usepackage{physics}
\usepackage{mathtools}
\usepackage{siunitx}

\usepackage{braket}
\usepackage{refcount}

\newlength{\singlecolumn}
\newcommand{\f}{\mathcal{F}}

\definecolor{mygreen}{RGB}{70, 170, 70}

\makeatletter
\renewcommand*{\@fnsymbol}[1]{\ensuremath{\ifcase#1\or \dagger\or *\or  \ddagger\or
\mathsection\or \mathparagraph\or \|\or **\or \dagger\dagger
\or \ddagger\ddagger \else\@ctrerr\fi}}
\makeatother

\newcommand{\appropto}{\mathrel{\vcenter{
  \offinterlineskip\halign{\hfil$##$\cr
    \propto\cr\noalign{\kern2pt}\sim\cr\noalign{\kern-2pt}}}}}

\def\app#1#2{
  \mathrel{
    \setbox0=\hbox{$#1\sim$}
    \setbox2=\hbox{
      \rlap{\hbox{$#1\propto$}}
      \lower1.1\ht0\box0
    }
    \raise0.25\ht2\box2
  }
}

\usepackage{hyperref}
\hypersetup{
    colorlinks,
    citecolor=black,
    filecolor=black,
    linkcolor=black,
    urlcolor=blue
}

\usepackage{graphicx}
\usepackage{changepage}

\usepackage{dirtytalk}
\usepackage{amstext}
\usepackage{array}  

\usepackage{amssymb}
\usepackage{textcomp}
\usepackage{bm}

\usepackage{mathtools}

\begin{document}

\title{Mixed Stochastic-Deterministic Approach for Many-Body Perturbation Theory Calculations}

\author{Aaron R. Altman}
\affiliation{Department of Materials Science and Engineering, Stanford University, Stanford, CA 94305, USA}

\author{Sudipta Kundu}
\affiliation{Department of Materials Science and Engineering, Stanford University, Stanford, CA 94305, USA}

\author{Felipe H. da Jornada}
\email{jornada@stanford.edu}
\affiliation{Department of Materials Science and Engineering, Stanford University, Stanford, CA 94305, USA}
\affiliation{Stanford Institute for Materials and Energy Sciences, SLAC National Accelerator Laboratory, Menlo Park, CA 94025, USA}

\begin{abstract} 
We present an approach for GW calculations of quasiparticle energies with \textit{quasi-quadratic} scaling by approximating high-energy contributions to the Green's function in its Lehmann representation with effective stochastic vectors. The method is easy to implement \textit{without altering the GW code}, converges rapidly with stochastic parameters, and treats systems of various dimensionality and screening response. Our calculations on a 5.75$^\circ$ twisted MoS$_2$ bilayer show how large-scale GW methods include geometry relaxations and electronic correlations on an equal basis in structurally nontrivial materials.
\end{abstract}

\maketitle

Many-body perturbation theory (MBPT) within the first-principles GW approximation is a proven and widespread method for computing accurate quasiparticle (QP) properties of materials~\cite{hedin1970effects, hybertsen1986electron, onida2002electronic, cohen2016fundamentals, louie2006conceptual}. Obtaining fully converged QP energies within the GW approach for small to moderately sized \textit{bulk} systems, with up to a few hundred atoms in the unit cell, is a routine procedure with modern high-performance supercomputers~\cite{sahni2020reliable, van2017automation, van2015gw, stuke2020atomic}. However, exploring the more complex many-body physics of large systems that are relevant in electronic and technological applications is difficult due to the quartic scaling in system size of the standard GW formalism~\cite{hybertsen1986electron}, limiting the applicability of the method in large-scale problems such those involving twisted materials displaying moir\'e physics~\cite{kennes2021simulator,naikPhysRevB.102.075413,wuPhysRevLett.122.086402,angeli2021gamma, carr2020electronic,tran2020moire,xian2021realization, lu2019modulated}.

Several approaches have been developed recently to deal with these shortcomings. They fall mainly into two categories: modifying the standard reciprocal-space GW formalism, wherein the electronic Green's function $G$ is still evaluated in its Lehmann representation as a sum-over-bands, or employing different representations of the theory that avoid the explicit sum-over-bands. Notable techniques in the former category include replacing high-energy orbitals with simple ansatz wavefunctions and using completion relations to truncate the sum-over-bands~\cite{samsonidze2011simple, gao2016speeding, bruneval2008accurate, deslippe2013coulomb, berger2010ab}.
Approaches in the latter category are diverse, with several achieving cubic or sub-cubic scaling. An important technique is transforming to bases where the evaluation of the polarizability is formally cubic scaling. This includes working in real space and imaginary time where the polarizability is separable~\cite{liu2016cubic, kim2020complex}, manipulating the spectral functions in a localized-orbital basis~\cite{foerster2011n}, exploiting sparse overlap integrals in a Gaussian basis~\cite{wilhelm2018toward}, 
and using tensor hypercontraction~\cite{parrish2013exact} and density-fitting 
methods~\cite{forster2020low, ma2021realizing, duchemin2021cubic}. Independent of these cubic scaling methods, stochastic approaches~\cite{neuhauser2014breaking, vlcek2017stochastic, vlvcek2018swift, romanova2022stochastic} can achieve linear scaling with system size by working in the time domain. Additionally, there are representations of the theory which still scale quarticly but exhibit lower prefactors, such as within the framework of density-functional perturbation theory~\cite{giustino2010gw,govoni2015large}.

In this letter, we propose a simple and rigorous approach that combines the stochastic and sum-over-bands methods to achieve a quasi-quadratic scaling GW formalism with a small prefactor (speedups of $\sim$100-fold on systems with tens of atoms), from given input mean-field wavefunctions. It offers large computational savings in both the calculation of the dielectric function and the QP self-energy. The performance gain is achieved by the stochastic compression of all mean-field Kohn-Sham states outside a small energy region around the Fermi level, \textit{including occupied states}.

Our approach is compatible with standard reciprocal-space GW codes and is simple to implement. It is also straightforward to converge independently of the GW code which uses it, and \emph{eliminates sum-over-bands truncation parameters in the GW calculation} by allowing one to include all eigenstates from the mean-field Hamiltonian. These advantages allow the computation of QP properties of complex systems of hundreds of atoms with moderate computational expense, which we demonstrate for several systems. Finally, unlike purely stochastic approaches, we observe speedups with respect to a fully deterministic approach for all system sizes, and not only for large systems. We highlight the applicability of our method on several systems of different dimensionality, including a large-scale problem of a 5.75$^\circ$ twisted bilayer of MoS$_2$.

\textit{Method.}---The GW approximation in its most common non-self-consistent form is based on the non-interacting single-particle Green's function,
\begin{align}
    G(\omega) &\equiv \sum_{n,{\vb{k}}} \frac{\ket{\phi_{n\vb{k}}}\bra{\phi_{n\vb{k}}}}{\omega - E_{n{\vb{k}}} \mp i\eta} \equiv \sum_{{\vb{k}}} G_{{\vb{k}}}(\omega)\label{green},
\end{align}
where $\ket{\phi_{n{\vb{k}}}}$ are mean-field states, typically obtained from density-functional theory (DFT) calculations, with band index $n$ and wavevector ${\vb{k}}$, $E_{n{\vb{k}}}$ are the corresponding eigenenergies, $\omega$ is the evaluation frequency, $\eta = 0^+$, and where the sign is negative (positive) when $E_{n{\vb{k}}}$ is below (above) the Fermi energy. As in other stochastic approaches to GW calculations~\cite{neuhauser2014breaking}, our method is based on the stochastic resolution of the identity operator,
$ \lim_{N\rightarrow \infty} N^{-1} \sum_{i=1}^N \ket{\zeta_i}\bra{\zeta_i} = \mathds{1}$, where off-diagonals vanish with a standard deviation of $1/\sqrt{N}$, and $\ket{\zeta_i}$ are random vectors (see Eq.~(\ref{SPBdef})).

When computing electronic properties within MBPT, it is important to accurately capture the pole structure of $G$ close to the Fermi energy. For instance, the non-interacting polarizability matrix $\chi^0_{\vb{G,G'}}(\vb{q},\omega)$ at a wavevector $\vb{q}$ and planewave indices $\vb{G}$ and $\vb{G}'$ has poles at frequencies corresponding to the energy difference between conduction ($c$) and valence ($v$) states, $\chi^0_{\vb{G,G'}}(\vb{q},\omega) \sim \sum_{vc\vb{k}} A^{vc}_{\vb{G,G'}}(\vb{k},\vb{q}) (\omega \pm (E_{c\vb{k}} - E_{v\vb{k}}))^{-1}$, where $A^{vc}_{\vb{G,G'}}$ are matrix elements. Accurately describing the \emph{low-frequency} behavior of $\chi^0$ is critical in MBPT calculations. This depends sensitively on the pole structure of $G$ close to the Fermi energy, but less so on the pole structure of $G$ at farther frequencies. For instance, when evaluating the electronic self-energy $\Sigma^{GW}$ within the contour-deformation approach~\cite{oschlies1995gw, lebegue2003implementation, bruneval2005exchange}, $\Sigma^{GW}$ depends on an integral of the screened Coulomb interaction $W$ along the imaginary frequency axis -- for which the pole structure of $\chi^0$ gets smoothed out -- plus residues of $W$ are typically evaluated at energies close to the Fermi energy.

\begin{figure}
    \centering
    \includegraphics[width=\singlecolumn,trim={.5cm .1cm .15cm .2cm},clip]{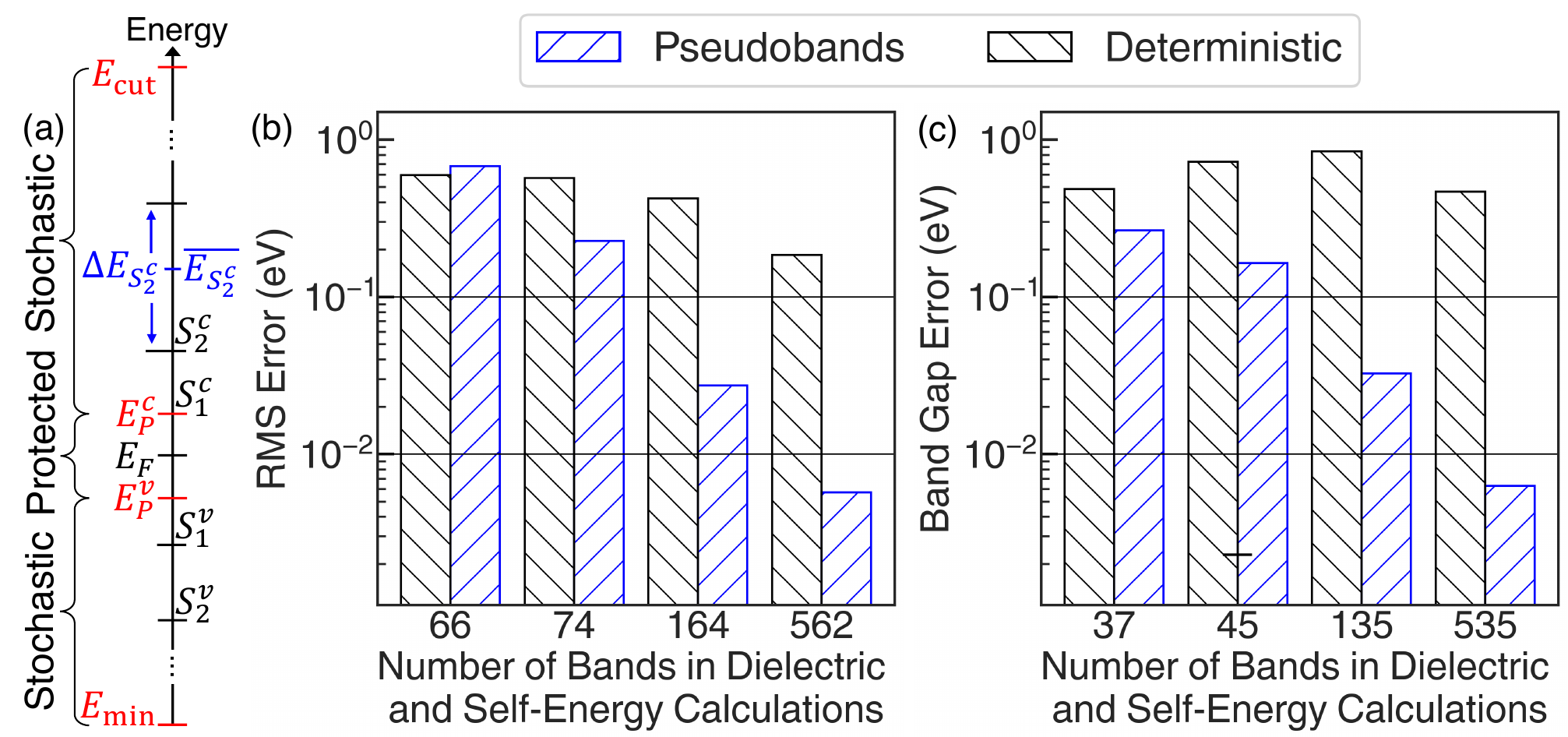}
    \caption{(a) Diagram of the method's band-partitioning scheme. $E_{\min}$ is the energy of the deepest valence state. (b/c) Comparison of the error in the QP energies for GW calculations performed with a traditional deterministic approach and using stochastic pseudobands for (b) an isolated benzene molecule and (c) bulk wurtzite ZnO. Stochastic pseudobands reach converged QP energies (within 10-100 meV) with fewer total bands than a deterministic truncation of the Hilbert space. Pseudobands parameters are (b) $N_P^c = 50$ and (c) $N_P^c = 10$; both (b) and (c) used $N_\xi^c = 2$ and $N_S^c = \{1,5,50,250\}$ (corresponding to $\f^c = \{1.9, 0.42, 0.054, 0.015\}$). Pseudobands were not used to compress valence states. }
    \label{benzene_ZnO}
\end{figure}

This motivates us to express $G$ as one term that contains the exact contributions to the pole structure close to the Fermi energy, $G_{{\vb{k}}}^P$, and another contribution that we write as a sum over $N_S$ subspaces that are farther from the Fermi energy, $G_{{\vb{k}}}^S$,
\begin{align}
    G_{{\vb{k}}}(\omega) &\approx G_{{\vb{k}}}^P(\omega) + \sum_S^{N_S} G_{{\vb{k}}}^S(\omega)\label{S-Dpart}.
\end{align}
$G_{{\vb{k}}}^P$ is computed exactly within Eq.~\ref{green} for bands $n \in P$, where $P$ is a \textit{small} protected subspace with $N_P$ bands closest to the Fermi energy. This deterministic region contains the states of interest for which QP properties are desired, though it is unnecessary when computing only the polarizability~\cite{SI_Npiszero}. The remaining subspaces $S$ are still required for accurately expressing the self-energy $\Sigma^{GW}$, but their pole structure may be approximated. For each subspace $S$, we first approximate the near-continuum pole distribution or branch cut at $\{E_{n\vb{k}}\}$ for states $n \in S$ with a single pole at an average energy $\bar{E}_{S\vb{k}}$. Next, we identify the sum $\sum_{n \in S} \ket{\phi_{n\vb{k}}}\bra{\phi_{n\vb{k}}}$ as a projection onto the subspace $S$. This projection can be compressed using the stochastic resolution of the identity operator,
\begin{align}
    G_{{\vb{k}}}^S(\omega) &\approx \frac{1}{\omega - \bar E_{S\vb{k}} \mp i\eta}\sum_{i=1}^{N_\xi}\ket{\xi^S_{i,{\vb{k}}}}\bra{\xi^S_{i,{\vb{k}}}},\label{SPBappx}
\end{align}
where $\ket{\xi^S_{i,\vb{k}}}$ are vectors that stochastically project any vector onto the subspace $S$ of interest, and which we denote by \emph{stochastic pseudobands}. Note that the subspaces $S$ can run over both unoccupied and occupied states [see Figure~\ref{benzene_ZnO}(a)].

The number of stochastic pseudobands $N_\xi$ is a convergence parameter and controls the stochastic error of the resolution of the identity. The number of subspaces $N_S$ is also a convergence parameter and controls the error of the average energy approximation. In the limit $N_S, N_\xi\rightarrow \infty$, we recover the original Green's function in Eq.~(\ref{green}). The partition Eq.~\eqref{S-Dpart} is a \emph{stochastic-deterministic} approach, and allows us to maintain high accuracy for important states close to the Fermi energy while compressing states that are less relevant. Our approach is similar in spirit to other stochastic methods for GW calculations~\cite{vlcek2017stochastic, vlvcek2018swift, romanova2022stochastic}, but does not require propagation in real-time.

Next, we show how to partition the subspaces $\{S\}$ in Eq.~(\ref{S-Dpart}) and construct each stochastic pseudoband $\ket{\xi^S_{i,{\vb{k}}}}$. A practical approach is to enforce that the error from each subspace to the Green's function or static polarizability matrix $\chi^0_{\vb{G,G'}}(\vb{q}, \omega=0)$ is roughly constant. This is achieved by enforcing a constant ratio
\begin{align}
    \f \equiv \dfrac{\Delta E_S}{\bar E_S} = \mathrm{const},\label{slicefrac}
\end{align}
where $\bar{E}_S$ is the average energy of the Kohn-Sham states in each subspace $S$ (referenced to the Fermi level) and $\Delta E_S$ is the energy range spanned by $S$~\cite{SIconvergence}. The ratio $\f$ is inversely proportional to the number of subspaces, $\f\sim 1/N_S$.

Finally, for each subspace $S$, we construct stochastic pseudobands by taking random linear combinations of Kohn-Sham states in $S$,
\begin{align}
    \ket{\xi^S_{i, \vb{k}}} &= \frac{1}{\sqrt{N_\xi}}\sum_{n\in S}\alpha^S_{i, n\vb{k}}\ket{\phi_{n\vb{k}}}\label{SPBdef},
\end{align}
with random phases $\alpha = e^{2\pi i \theta}$ for random $\theta \in [0,1)$, and $i\in \{1,\dots, N_\xi\}$ are the different stochastic pseudobands that realize the projection onto $S$.

The proposed stochastic compression can be easily implemented in most MBPT codes that use a spectral representation of $G$, and we have implemented our developmental version in the BerkeleyGW code~\cite{deslippe2012berkeleygw}. One only needs to modify the input Kohn-Sham orbitals and combine them according to Eq.~\eqref{SPBdef}. In particular, no modification of the GW code is required: the pseudobands approach is a pre-processing step to the GW calculation. We also provide a pseudocode~\cite{SI_pseudocode} and reference implementation~\cite{altman_Zenodo}. The method as described here focuses on compressing the Green's function for the efficient evaluation of the static dielectric function, which is the quantity of interest in calculations that use plasmon-pole models. Still, a simple extension, whereby one takes $\Delta E_S$ to be a constant ($\Delta E$) instead of a quantity proportional to $\bar{E}_S$, allows the evaluation of the inverse dielectric function at arbitrary frequencies with small statistical errors and large computational savings~\cite{SIcompdeets}. We stress that our approach is amenable to compressing both valence and conduction states, offering especially large speedups for the computation of the dielectric function, which scales with their product. We summarize the quantities introduced in Table~\ref{tab:params} below.

\begin{table}[H]
\caption{\label{tab:params} Pseudobands parameters: `conv.', `auto.', and `aux.' are convergence, automatically determined, and auxiliary parameters, respectively. $N_P$ should be 0 when evaluating the polarizability for large systems and finite to evaluate the GW self-energy of deterministic states.}
\begin{ruledtabular}
\begin{tabular}{r@{}lcccr@{}l}
\multicolumn{2}{c}{Parameter} & & Description & & \multicolumn{2}{c}{\hspace{-18pt}Typical Value} \\
\hline
$\f$ &-- conv. & & Constant energy ratio Eq.~(\ref{slicefrac})  & & 1\% \ &-- 2\% \hspace{.5mm} \\
$N_S$ &-- auto. & & Number of stochastic subspaces, $N_S\appropto\frac{1}{\f}$ & & 10 \ &-- 200 \\
$N_\xi$ &-- conv. & & Number of pseudobands per subspace & & 2 \ &-- 3 \ \ \\
$N_P$ &-- aux. & & Number of protected bands & & \ &$\geq$0 \  \\
\end{tabular}
\end{ruledtabular}
\end{table}

We note that (1) convergence testing with respect to pseudobands parameters is rarely required, as the typical values listed in Table~\ref{tab:params} were sufficient to converge all systems studied and (2) $N_P$ only needs to be large enough to include the states of interest for computing the electronic self-energy, and can be zero for computing only the dielectric matrix for both semiconductors and metals~\cite{SI_Npiszero}. Additionally, our approach \emph{removes the band truncation parameters} employed in the sum-over-states in traditional GW calculations. This is because, when constructing the stochastic pseudobands, we can easily consider all bands from the mean-field Hamiltonian by diagonalizing it with scalable linear algebra packages such as ELPA~\cite{marek2014elpa}. While one can benefit from similar speedups from our pseudobands approach when generating input Kohn-Sham states with iterative solvers, directly diagonalizing the DFT Hamiltonian is typically faster and more numerically stable~\cite{SImfham}.

\textit{Results.}---We benchmark our stochastic pseudobands approach on systems spanning dimensionality, electronic structure, and screening environment to numerically verify its convergence behavior. We demonstrate quasi-quadratic scaling for GW calculations on ZnO supercells up to 256 atoms while maintaining constant error. Finally, we perform a large-scale calculation of the GW QP bandstructure of a 5.75$^\circ$ twisted MoS$_{2}$ moir\'e bilayer to address questions regarding the emergent electronic structure in twisted 2D materials. Computational details are provided in the Supplemental Material~\cite{SIcompdeets}. Specific pseudobands convergence parameters are listed with the computations below ($v/c$ superscripts indicate pseudobands parameters used for valence/conduction states, respectively). Regardless of the system, we note that $N_\xi \geq 2$ should be used, as $N_\xi = 1$ does not resolve the projection over each subspace. Additionally, as currently implemented, compressing \emph{valence} states with stochastic pseudobands does not offer advantages for calculating the self-energy operator (as opposed to the dielectric matrix). This is because the bare exchange contribution to the self-energy $\Sigma^X$, which involves matrix elements with occupied states, is very sensitive to the character of the valence wavefunctions. Since the calculation of the self-energy only scales with the \emph{sum} of the valence and conduction bands, compressing valence states does not provide significant acceleration for $\Sigma^{GW}$ in any case. However, stochastic pseudobands always provide speedups when compressing the conduction states for the operators studied here.

\textit{Convergence Behavior.}---We show systematic convergence of QP energies for two systems, an isolated benzene molecule and bulk wurtzite ZnO. Additional benchmarks on bilayer MoS$_2$ and a metallic Ag$_{54}$Pd nanoparticle are presented in the Supplemental Material~\cite{SIextraplots}. Figure~\ref{benzene_ZnO} summarizes our approach by comparing the error in QP energies for an isolated benzene molecule and for wurtzite ZnO with respect to the number of bands $N_b$ included in the MBPT calculations -- both in the summations to evaluate the dielectric matrix and self-energy. For each value of $N_b$, we include either the lowest $N_b$ Kohn-Sham orbitals, in the deterministic case, or both a set of $N_P$ Kohn-Sham states in the protected region plus $N_S N_\xi$ stochastic pseudobands, such that $N_b = N_P + N_S N_\xi$. Hence, our tests assess whether, for a fixed computational effort, stochastic pseudobands yields more accurate QP energies by approximating the high-energy part of the Hilbert space that gets truncated in deterministic calculations. Figure~\ref{benzene_ZnO}(b) shows the root mean square (RMS) error $\sqrt{\frac{1}{N}\sum_n^N\left(E^{\mathrm{QP}}_n - E^\mathrm{ref}_n\right)^2}$ over 19 QP levels around the Fermi energy of benzene for both the deterministic calculation and pseudobands. Figure~\ref{benzene_ZnO}(c) shows the error of the bandgap $\left|E_{\mathrm{gap}}^{\mathrm{QP}} - E^\mathrm{ref}_{\mathrm{gap}}\right|$ of ZnO, again comparing both the deterministic calculations and those using stochastic pseudobands. In both cases, $E^\mathrm{ref}$ is obtained from a highly converged deterministic calculation -- utilizing 30,000 bands for benzene and 10,000 bands for ZnO.

For both materials, stochastic pseudobands outperforms the deterministic results by 10-100-fold in error for the same computational effort for all but the least converged calculations. Conversely, we find that, to achieve the same error, the deterministic calculation requires approximately 10-100 times as many bands as used in stochastic pseudobands calculations. We see rapid and systematic convergence behavior for all systems studied.

\begin{figure}
    \centering
    \includegraphics[width=\singlecolumn, trim={.4cm .5cm 2cm .5cm},clip]{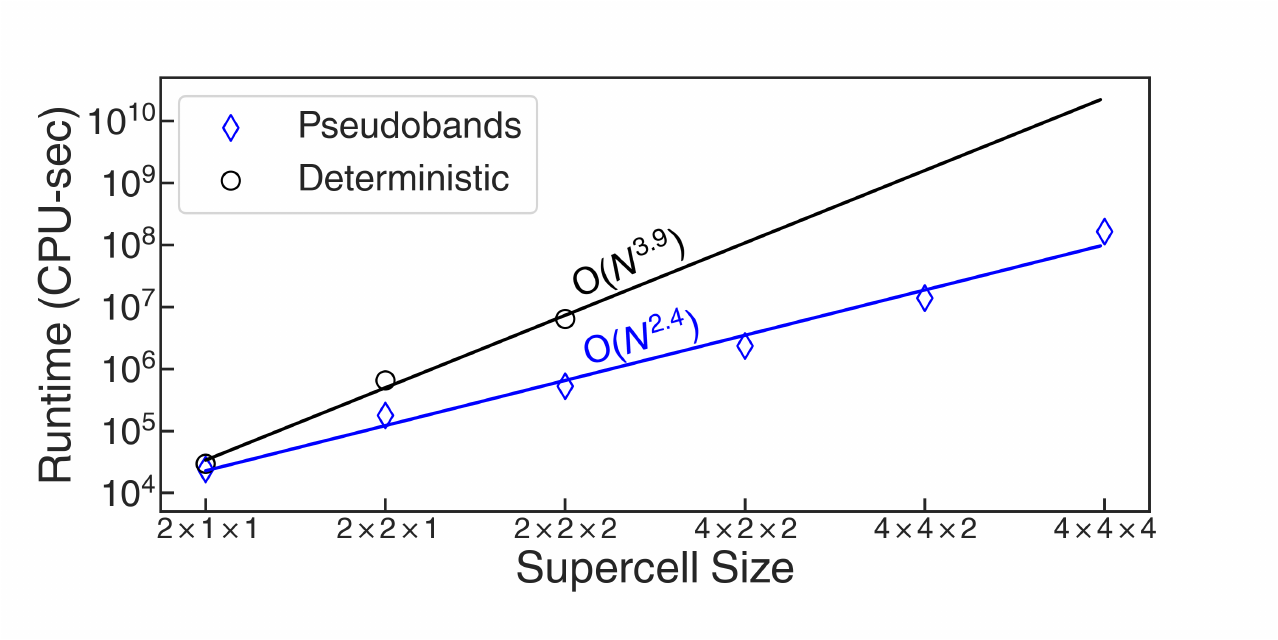}
    \caption{Scaling curve for the dielectric computation per q/k-point for ZnO supercells showing quasi-quadratic behavior up to 256 atoms. Bandgap errors are maintained at $<50$ meV for constant convergence parameters $N_\xi, \f$~\cite{SIextraplots}.}
    \label{eps_scaling}
\end{figure}

\textit{Scaling and Computational Cost.}---In addition to the good convergence behavior, utilization of stochastic pseudobands also significantly improves the computational scaling of the GW approach with system size. Traditionally, the calculation of the dielectric matrix consists of two primary computationally demanding steps: constructing the non-interacting polarizability matrix $\chi^0_{\vb{G,G'}}$, which scales as $\mathcal{O}(N_{\vb{G}}^2\times N_c\times N_v) \sim \mathcal{O}(N^4)$, and then inverting the RPA dielectric matrix $\epsilon_{\vb{G,G'}}$, which scales as $\mathcal{O}(N_{\vb{G}}^3) \sim \mathcal{O}(N^3)$, where $N_v$ and $N_c$ are the numbers of valence and conduction bands, and $N_{\vb{G}}$ and $N$ are the number of reciprocal-lattice vectors and the overall system size, respectively. With stochastic pseudobands, the cost to compute the noninteracting polarizability is $\mathcal{O}(N_G^2N_S^2N_\xi^2)$, since one can always take $N_P=0$. It still takes $\mathcal{O}(N^3)$ to invert the dielectric matrix, although that cost can be reduced with low-rank techniques~\cite{wilson2008efficient, shao2016low, del2019static}, making the GW workflow quasi-quadratic. Our results show that the computational savings are insensitive to the details of $N_P$ (see Figure~\ref{eps_scaling}). Additionally,  from Eq.~(\ref{slicefrac}), the total number of states when utilizing pseudobands is roughly the logarithm of the initial number of states, yielding a significant reduction in the number of states used in the MBPT calculations and a low algorithmic prefactor. Moreover, due to the high performance of distributed-memory linear algebra solvers, we find that the inversion of the dielectric matrix is only a significant bottleneck for large systems, with hundreds to thousands of atoms in the unit cell. In fact, for the largest system we studied of 5.75$^\circ$ twisted bilayer MoS$_2$, inversion took only 14\% of the total run time. 

Figure~\ref{eps_scaling} shows the computational scaling for calculating a well-converged dielectric matrix for ZnO, with a plane-wave cutoff of 80~Ry, where we consider systematically larger supercells containing from 8 to 256 atoms~\cite{SIcompdeets}. To make these calculations feasible, it was critical to use our approach wherein both valence and conduction states away from the Fermi energy are compressed into stochastic pseudobands. All supercells exhibited constant error $< 50$~meV when we performed subsequent self-energy calculations of the quasiparticle energies with unchanged convergence parameters ($N^v_\xi = 4$, $N_\xi^c = 2$, $\f^{v/c} = 0.02$). $N_P^{v/c}$ was chosen to be 20 for the $2\times 1\times 1$ supercell and scales with the system size to allow for the evaluation of the self-energy within the same energy window~\cite{SIcompdeets, SIextraplots, SI_Npiszero}). Even with a nonzero $N_P$, we find that the approach displays in practice a \emph{quasi-quadratic scaling} for the evaluation of the dielectric function and self-energy.

\textit{Application to Large Systems.}--- Moir\'e bilayers such as twisted bilayer graphene or transition metal dichalcogenides (TMDs) have been at the research forefront for investigating correlated electronic phases in condensed matter systems ~\cite{andrei2021marvels, zheng2020unconventional, cao2018graphene, wang2020wse2,kennes2021simulator,xu2020correlated}. Semiconducting TMD moir\'e bilayers have gained additional interest as hosts of different types of emergent excitons for possible applications in optoelectronic and exciton-based qubit devices~\cite{tran2019evidence, yu2017moire,andersen2021excitons,naik2022intralayer,jauregui2019electrical,karni2019infrared,karni2022structure,tang2021tuning,seyler2019signatures,shimazaki2020strongly,jin2018imaging}. A correct description of the QP properties is often a prerequisite to understanding these emergent phenomena, but the large system size and variations of the dielectric function~\cite{Latini2015,Qiu2016} requiring fine Brillouin zone (BZ) samplings~\cite{daJornada2017} has made them difficult to study with first-principles GW calculations.

\begin{figure}
    \centering   
    \includegraphics[width=\singlecolumn, trim={.2cm 0cm 0cm 0.0cm},clip]{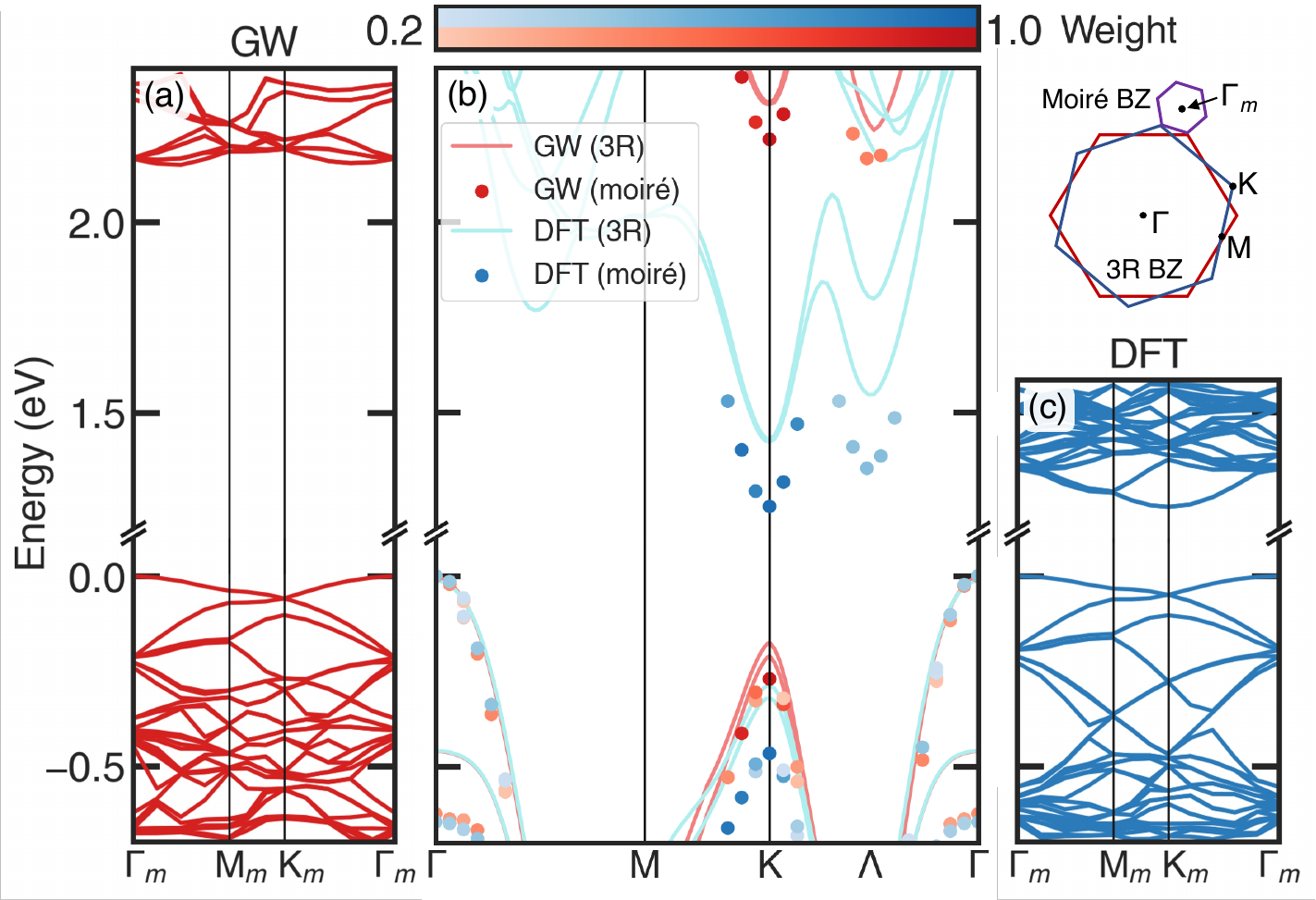}
    \caption{(a/c) GW (computed with stochastic pseudobands) and DFT band structure for the 5.75$^\circ$-twisted moiré system in the moiré BZ. (b) Band structure for the untwisted, 3R-stacked MoS$_2$ bilayer (solid lines) and the corresponding unfolded band structure of the $5.75^\circ$ twisted moir\'e system (dots), at the DFT and GW levels. The weight of the projection represents the contribution of the unit cell state to the corresponding moir\'e state at the same energy. Inset above (c): BZs of the twisted and untwisted structures.}
    \label{bigMoS2}
\end{figure}

Using the pseudobands approach, we perform explicit GW calculations on a 5.75$^\circ$-twisted bilayer of MoS$_2$ containing 546 atoms in the moiré supercell [Figure~\ref{bigMoS2}(a)], and further unfold the moiré band structure to the unit cell of the high-symmetry, 0$^\circ$-twisted bilayer structure~\cite{kundu2022exciton, popescu2012extracting}, known as the 3R stacking [red dots in Figure~\ref{bigMoS2}(b)]. We compare such large-scale GW calculations to DFT calculations performed directly on the 3R structure [blue lines in Figure~\ref{bigMoS2}(b)] and find differences in the bandgap, the relative energy splitting, and band ordering.

We rationalize these differences through contributions from moir\'e and quasiparticle effects. To understand moir\'e effects, we perform DFT calculations on the twisted structure [Figure~\ref{bigMoS2}(c)] and unfold the resulting band structure onto the BZ of the 3R structure [blue dots in Figure~\ref{bigMoS2}(b)]. Compared to direct DFT calculations on the 3R structure, the DFT calculations on the 5.75$^{\circ}$-twisted system display a larger energy splitting between the first two valence states at $\Gamma$. This is expected since these states originate from the interlayer chalcogen interactions, which change with stacking and twist angle. Next, to capture quasiparticle effects, we perform GW calculations on the 3R structure [red lines in Figure~\ref{bigMoS2}(b)]. Compared again to the DFT calculation on the 3R structure, QP effects mainly increase the bandgap and reorder the conduction states at the $K$ and $\Lambda$ valleys. Our calculations highlight that moir\'e effects and quasiparticle corrections both play significant roles in twisted materials and need to be accounted for on the same footing~\cite{lu2019modulated}, and are additive here to within 100 meV~\cite{SI_moire_extrapolate}.

\textit{Conclusion.}---We present a mixed stochastic-deterministic approach for GW calculations. Given input mean-field states, the method displays quasi-quadratic scaling for the tests performed up to 256 atoms, $\sim$100-fold speedups for systems of tens of atoms, and smooth convergence behavior. The usage of stochastic pseudobands is compatible with systems of any dimension, nontrivial screening environments, and extends standard MBPT codes to handle systems of several hundreds of atoms with moderate computational expense. We envision that, beyond further studies on moiré systems, structurally large and technologically relevant systems such as interfaces, surfaces, and extended defects can be studied with this approach to address fundamental questions involving the interplay between nonlocal screening environments and self-energy effects.

\begin{acknowledgments}
ARA acknowledges helpful discussions with Mauro Del Ben, Johnathan D. Georgaras, and Emma M. Simmerman. This work was primarily supported by the Center for Computational Study of Excited-State Phenomena in Energy Materials (C2SEPEM), funded by the U.S. Department of Energy (DOE), Office of Basic Energy Sciences (BES) under Contract No. DE-AC02-05CH11231 at the Lawrence Berkeley National Laboratory (LBL), as part of the Computational Materials Sciences Program. The study of twisted MoS$_2$ was supported by the U.S. DOE BES grant DE-SC0021984. This research used resources of the National Energy Research Scientific Computing Center (NERSC), a U.S. DOE Office of Science User Facility located at LBL, operated under Contract No. DE-AC02-05CH11231 using NERSC award BES-ERCAP m3606, and from the Texas Advanced Computing Center (TACC) at The University of Texas at Austin, funded by the National Science Foundation (NSF) award 1818253, through allocation DMR21077 for the development of algorithms. Large-scale calculations used resources of the Oak Ridge Leadership Computing Facility at the Oak Ridge National Laboratory, which is supported by the U.S. DOE BES under Contract No. DE-AC05-00OR22725.
\end{acknowledgments}

\bibliography{bibliography}

\end{document}


\title{Supplementary Material for Mixed Stochastic-Deterministic Approach for Many-Body Perturbation Theory Calculations}
\author[1]{Aaron R. Altman}
\author[1]{Sudipta Kundu}
\author[1,2]{Felipe H. da Jornada\footnote{E-mail: jornada@stanford.edu}}
\affil[1]{Department of Materials Science and Engineering, Stanford University, Stanford, CA 94305, USA}
\affil[2]{Stanford Institute for Materials and Energy Sciences, SLAC National Accelerator Laboratory, Menlo Park, CA 94025, USA}

\date{}

\maketitle
\tableofcontents

\section{Computational Details}\label{compdeets}
Subsections below include material-specific details such as convergence parameters. See Ref.~\cite{altman_Zenodo} for the atomic structures. For all materials, density-functional theory (DFT) calculations were performed within the Quantum ESPRESSO package~\cite{giannozzi2009quantum} with the Perdew-Burke-Ernzerhof (PBE) exchange-correlation functional~\cite{perdew1996generalized} to obtain mean-field energies and wavefunctions. All calculations used the plane-wave basis and scalar-relativistic norm-conserving pseudopotentials~\cite{van2018pseudodojo, schlipf2015optimization}. GW calculations were performed in the BerkeleyGW code~\cite{deslippe2012berkeleygw} in a single-shot approach. Reference GW calculations used for comparison with stochastic pseudobands GW calculations utilized enough unoccupied states in the sum-over-bands to be effectively fully converged with respect to the other cutoffs used, as we detail below for each system.

For details about strong and weak scaling of the BerkeleyGW code, we refer the reader to Ref~\cite{deslippe2012berkeleygw}.

In section S\ref{exactdiag} we describe our approach to obtaining the large numbers of unoccupied mean-field states required for GW calculations. In section S\ref{pseudocode} we provide pseudocode and reference to standalone Python scripts for our implementation of pseudobands. The scripts are agnostic to the input mean-field wavefunctions and the GW code which utilizes the output wavefunctions, except for file format. 

\subsection{Benzene}
Benzene atomic structure is taken from the GW100 benchmark~\cite{van2015gw}. The wavefunction plane-wave cutoff used was 50 Ry, a unit cell that encloses eight times the volume containing 95\% of the mean-field charge density was used, and 30,000 mean-field states were generated. We chose a modest unit cell and energy cutoffs to allow us to perform fully converged deterministic calculations without relying on extrapolations of convergence parameters. The GW calculations employed a 12 Ry dielectric cutoff, the Hybertsen-Louie Generalized Plasmon-Pole (GPP) model for the frequency dependence of the dielectric function~\cite{hybertsen1986electron}, and a box-truncation scheme of the Coulomb interaction to avoid spurious effects from periodic images~\cite{ismail2006truncation}. The self-energy was evaluated for the 10 highest occupied and 9 lowest unoccupied molecular orbitals. For the reference GW calculation, \emph{all} 30,000 Kohn-Sham states were used (i.e., all the Kohn-Sham states available given the simulation unit cell and wavefunction cutoff). Pseudobands compressed the same 30,000 states into varying numbers of stochastic bands. Root-mean-square (RMS) quasiparticle (QP) errors when using pseudobands for GW were computed over these 19 states. 

The stochastic pseudobands parameters used were $N_P^c = 50$, $N_\xi^c = 2$, $N_S^c = \{1,5,50,250\}$. $N_P=50$ is enough to capture all relevant band reordering. We did not employ stochastic pseudobands to compress valence states since a single benzene molecule has a small number of occupied orbitals.

\subsection{ZnO}
ZnO crystal structure is taken from experiment, obtained from the American Mineralogist Crystal Structure Database (\_database\_code\_amcsd 0005203)~\cite{downs2003american}. A wavefunction cutoff of 200 Ry was used with a $4\times 4\times 4$ k-grid, and 10,000 mean-field states were generated. For the GW calculation, a dielectric cutoff of 80 Ry was used with frequency dependence from the GPP model, and the self-energy was evaluated at the $\Gamma$ point to compute the valence band maximum (VBM) and conduction band minimum (CBM). With these parameters we obtain a band gap of 2.69 eV -- within 100 meV of a recently reported value~\cite{rangel2020reproducibility}.

Pseudobands parameters used were $N_P^c = 10$, $N_\xi^c = 2$, $N_S^c = \{1,5,50,250\}$. Valence pseudobands were not used.

Supercell structures for the scaling test were generated with the Python module \texttt{ase}~\cite{larsen2017atomic}. Deterministic GW calculations were performed for the $2\times 1\times 1$, $2\times 2\times 1$, and $2\times 2\times 2$ supercells to verify stability of the QP gap with respect to supercell size -- the deviation from the unit cell GW gap was less than 1 meV for different supercell sizes (variations in the GW gap are related to the numerical treatment of the $q\rightarrow 0$ limit of the screened Coulomb interaction). Thus, all errors from utilizing stochastic pseudobands were measured with respect to the unit-cell QP values.

We chose the number of protected bands to scale with the supercell size to account for the effects of band-folding and avoid an increase in the error in the QP energies. We picked $N_P^{v/c}$ being 10 times the number of ZnO unit cells in each supercell calculation. The other parameters were fixed at $N_\xi^v = 4$, $N_\xi^c = 2$, $N_S^v = 5$, and $N_S^c = 185 \pm 10$. There were slight variations of $N_S^c$ due to differences in how the bands fell into the allocated slices depending on the supercell size. All runs used the slice ratio $\f = 0.02$.

\subsection{MoS${}_2$}\label{MoS2details}
The AA stacked bilayer MoS${}_2$ structure was obtained with relaxation in the LAMMPS code~\cite{thompson2022lammps} with the Kolmogorov-Crespi interlayer potential~\cite{naik2019kolmogorov} and Stillinger–Weber intralayer potential~\cite{jiang2013molecular}. The $5.75^\circ$ twisted MoS${}_2$ bilayer was generated with a homemade moire-structure code and relaxed in LAMMPS with the same potentials.

For the AA bilayer, we used a wavefunction cutoff of 35 Ry, the unit cell extended 20 \AA \ in the non-periodic direction, and a $6\times 6\times 1$ k-grid. Convergence tests of all relevant direct and indirect QP valence-conduction gaps (e.g. $K-\Gamma$, $\Gamma-K$, $K-K$, $M-K$, etc.) with respect to the wavefunction cutoff were performed, and the 35 Ry cutoff was sufficient to converge these gaps to within 30 meV with of the QP gaps from a 70 Ry wavefunction cutoff calculation. To properly converge the dielectric function we employed the nonuniform neck subsampling (NNS) technique~\cite{felipe2017nonuniform} for q-points near $\Gamma$. We included 4,055 mean-field states for the GW calculations~\cite{qiu2013optical}, truncated the Coulomb interaction along the non-periodic direction, and employed a cutoff of 35 Ry for the dielectric matrix with frequency dependence from the GPP model. The self-energy was evaluated for states VBM-4 through CBM+4 with a 15 Ry cutoff for the screened Coulomb interaction, enough to converge relative QP energies at the K and $\Lambda_{\mathrm{min}}$ points of the lowest-energy conduction bands to within about $100$~meV.

Extensive convergence testing for pseudobands was not performed for this system. Instead, we chose the following reasonable parameters and obtained excellent results for the QP energies relative to the deterministic calculation: $N_P^c = 10, N_\xi^c = 3, N_S^c = 151$ (the corresponding slice fraction $\f = 0.02$). Valence pseudobands were not used due to the small number of occupied states.

For the $5.75^\circ$ twisted bilayer, we could only perform the calculation with stochastic pseudobands. We employed a $3\times 3\times 1$ k-grid with a 35 Ry wavefunction cutoff at the mean-field level, and a 20~\AA\ cell in the non-periodic direction. For the dielectric function we used a 25 Ry cutoff, and the GPP model was used for the frequency dependence. NNS was not employed as the $3\times 3\times 1$ q-grid is fine enough to sample the sharp peak in the dielectric function due to the large cell size. The self-energy was computed with a 15 Ry cutoff for the screened Coulomb interaction for 27 valence and 13 conduction states, and linear interpolation in the moir\'e BZ gave the QP band structure. Pseudobands compressed all states up to the wavefunction cutoff (380,000 input Kohn-Sham orbitals were used to create all stochastic pseudobands). The parameters used were $N_P^{v/c} = 1000, N_\xi^{v/c} = 3, N_S^v = 36, N_S^c = 211$, corresponding to $\f^{v/c}=0.02$. This value of $N_P$ corresponds to the AA-stacked bilayer scaled by the supercell size ($\sim 10\times 10$).

\subsection{Ag${}_{54}$Pd}\label{AgNpdetails}
An Ag${}_{55}$ icosahedral nanoparticle structure was generated with the \texttt{ase} Python package~\cite{larsen2017atomic}. 
One of the Ag atoms on an edge of the icosahedron was then replaced with a Pd atom, without relaxation, to make Ag${}_{54}$Pd. A 25~\AA${}^3$ cubic unit cell was used, with a somewhat small wavefunction cutoff of 23 Ry to demonstrate the method. We generated 10,000 mean-field states. Rather than using the GPP model, the dielectric function was computed explicitly in frequency space from 2.5 eV to 4.5 eV with a 100 meV grid spacing and a 200 meV broadening. We employed the static subspace approximation to speed up the full-frequency calculation with a basis size of 2500 eigenstates~\cite{del2019static}. The self-energy was not computed; convergence of pseudobands was measured with respect to the macroscopic dielectric function $\epsilon^M(\omega) = 1/\epsilon^{-1}_{00}(\omega)$. 
 
To make the stochastic pseudobands approach amenable to evaluating the dielectric matrix at arbitrary frequencies, we slightly modify the way energy subspaces are chosen (see section S\ref{fullfreq} below). In short, we include a low-energy region up to $\omega_\max$ where the energy spanned by each subspace $\Delta E_S = \delta\omega$ is constant, after which exponential slices begin as usual. The following pseudobands parameters were used: $N_P^{v/c} = 10, N_\xi^{v/c} = \{1,5\}, N_S^v = \{2, 10, 20, 50\}, N_S^c = \{2, 10, 100, 500\}$ (corresponding to $\f^v = \{0.51, 
0.11, 0.11, 0.0079\}$ and $\f^c = \{0.67, 0.14, 0.014, 0.0027\}$). Here the $N_S$ parameters only describe the exponential portion of the slices. $\omega_\max$ and $\delta\omega$ were determined by the energy range and spectral resolution of the desired calculation of the dielectric matrix; we set $\omega_\max = 10.2$~eV and $\delta\omega = 100$~meV. In principle $\omega_{\mathrm{max}}$ can be made smaller, and convergence testing should be performed for real calculations.

\subsection{Obtaining Unoccupied Mean-Field States}\label{exactdiag}
In this subsection, we briefly clarify our approach to generating the large number of unoccupied states required for the GW calculations, which is independent of the pseudobands approach. \emph{We emphasize that while the approach described below is very efficient, it is not required to apply the pseudobands method, which is agnostic to the procedure used to generate the mean-field states.}  

As described in the main text, we perform a direct diagonalization of the mean-field Hamiltonian $H^{\mathrm{MF}}$ with a highly optimized linear algebra solver such as ELPA~\cite{marek2014elpa} to obtain all mean-field states up to the chosen plane-wave cutoff. When the number of bands is a significant fraction ($>$10\%) of the size of the Hamiltonian, iterative diagonalizers become inefficient and/or numerically unstable. Therefore, there are at least three clear advantages in diagonalizing the full Kohn-Sham Hamiltonian as opposed to using an iterative solver:
\begin{enumerate}
\item \underline{Well-converged wavefunctions}: iterative diagonalizers in DFT codes are optimized for the low-energy part of the energy spectrum. When solving for a large fraction of the eigenspace, they are inefficient and often struggle to obtain well-converged eigenvectors, especially when one requests thousands of Kohn-Sham states. Poorly converged states can lead to an unphysically large screening response, and is easily avoided with a direct diagonalizer. 
    
\item \underline{Easier convergence}: a direct matrix diagonalization algorithm yields the whole unoccupied manifold described up to the wavefunction cutoff, facilitating convergence testing. There is no concern about having to recompute unoccupied Kohn-Sham states if the previous amount was insufficient. Additionally, when coupled with the pseudobands approach, a direct diagonalization of the DFT Hamiltonian further eliminates the convergence parameters of how many bands to include: they are all included within our stochastic formalism.

\item \underline{Significant speedup}: a major benefit of directly diagonalizing the Kohn-Sham Hamiltonian is that, in practice, it is significantly faster than using iterative diagonalizers when one is interested in more than about 5\% of the spectrum. For reference, below we include previous benchmarks we performed on Edison at the National Energy Research Scientific Computing Center (NERSC), comparing a direct diagonalizer, Parabands, bundled with the BerkeleyGW package~\cite{deslippe2012berkeleygw} and using the ELPA library, as well as reference values of the Quantum ESPRESSO~\cite{giannozzi2009quantum} Davidson algorithm.
    \begin{center}
\begin{table}[H]
\def\arraystretch{1.3}
\caption{\label{tab:parabands} Benchmarks for Direct vs. Iterative Diagonalization of $H^{\mathrm{MF}}$}
\begin{tabular}{ | m{3.8cm} | m{3.5cm}| m{3.5cm} | m{3.5cm}| } 
  \hline
  System & N${}_2$ molecule & TTF molecule & MoS${}_2$ monolayer\\ 
  \hline
  Size of Hamiltonian $H^{\mathrm{MF}}$ & 36k & 137k & 77k\\ 
  \hline
  Number of bands generated & 6.5k (18\% of $H^{\mathrm{MF}}$) & 27k (20\% of $H^{\mathrm{MF}}$) & 5.1k (7\% of $H^{\mathrm{MF}}$)\\ 
  \hline
  Wall time: Quantum ESPRESSO (Davidson) & 3.5h \newline(256 tasks $\times$ 4 threads) & $>$4h (never finished)\newline(512 tasks $\times$ 1 thread) & 3.5h \newline (64 tasks $\times$ 1 thread)\\ 
  \hline
  Wall time: Parabands (ELPA 1-stage) & 4 mins \newline (256 tasks $\times$ 1 thread) & 26 mins \newline (512 tasks $\times$ 1 thread) & 11 mins \newline (512 tasks $\times$ 4 threads) \\ 
  \hline
\end{tabular}
\end{table}
\end{center}

\end{enumerate}

We also briefly describe the implementation of the exact diagonalization of $H^{\mathrm{MF}}$ within the Parabands code in BerkeleyGW: Parabands reads as input the Kohn-Sham potential and wavefunctions obtained from a self-consistent-field calculation of a DFT code. Then, it uses the non-local parts of the pseudopotentials in Kleinman-Bylander form to construct $H^{\mathrm{MF}}$ on-the-fly in the plane-wave basis. Finally, a direct diagonalizer such as ELPA is called, and the resulting wavefunctions and eigenenergies are written to file.

Although the exact diagonalization approach is very efficient for getting the large numbers of unoccupied states required to converge GW sums-over-bands, the pseudobands approach will provide the same benefits regardless of how these states are generated, as long as there are enough to otherwise converge the GW calculations.

Finally, we note that, for all systems studied, the diagonalization of the mean-field Hamiltonian took less time than the evaluation of the dielectric matrix when utilizing the pseudobands approach. Even for the largest ZnO supercell studied, the diagonalization of the Kohn-Sham Hamiltonian took only 38\% of the computational resources necessary to evaluate the dielectric matrix, and for the  MoS$_2$ moiré system, the diagonalization took 88\% of the resources of the dielectric calculation.

\subsection{Pseudocode and Python Scripts for Pseudobands}\label{pseudocode}

We present a pseudocode for our implementation of pseudobands, and provide a reference to a standalone Python implementation~\cite{altman_Zenodo}. The Python scripts are agnostic to the mean-field code that generates the input mean-field states and to the GW code which uses the output. The only assumption is that the input wavefunctions are in BerkeleyGW format\footnote{\url{http://manual.berkeleygw.org/3.0/wfn_h5_spec/}}. The pseudocode follows:

\begin{algorithm}
\SetAlgoLined
\DontPrintSemicolon
\SetKwData{wfn}{wfn}\SetKwData{wfnout}{wfn\_out}\SetKwData{nk}{n\_k}\SetKwData{nb}{n\_b}\SetKwData{nspb}{n\_spb}\SetKwData{nbasis}{n\_basis}\SetKwData{en}{en}\SetKwData{enout}{en\_out}\SetKwData{env}{en\_v}\SetKwData{envtmp}{en\_v\_tmp}\SetKwData{enc}{en\_c}\SetKwData{first}{first}\SetKwData{last}{last}
\SetKwData{firstidx}{first\_idx}\SetKwData{lastidx}{last\_idx}
\SetKwData{slicec}{slices\_c}\SetKwData{slice}{slice}\SetKwData{slicev}{slices\_v}
\SetKwData{nsc}{ns\_c}\SetKwData{npc}{np\_c}\SetKwData{nxc}{n$\xi$\_c}\SetKwData{fc}{$\f$\_c}
\SetKwData{nsv}{ns\_v}\SetKwData{npv}{np\_v}\SetKwData{nxv}{n$\xi$\_v}\SetKwData{fv}{$\f$\_v}
\SetKwData{complex}{\textbf{complex}}\SetKwData{float}{\textbf{float}}\SetKwData{int}{\textbf{int}}
\SetKwData{ik}{ik}
\SetKwData{shift}{shift}
\SetKwData{sliceidx}{slice\_idx}
\SetKwData{ix}{i$\xi$}
\SetKwData{phases}{phases}
\SetKwData{nbs}{n\_b\_slice}
\SetKwInOut{Input}{input}\SetKwInOut{Output}{output}
\SetKwFunction{min}{min}
\SetKwFunction{len}{len}
\SetKwFunction{mysqrt}{sqrt}
\SetKwFunction{mean}{mean}
\SetKwFunction{range}{range}
\SetKwFunction{matvec}{matvec}
\SetKwFunction{findbandidx}{find\_band\_idx}
\SetKwFunction{append}{append}
\SetKwFunction{constructslices}{construct\_slices}
\SetKwFunction{constructspb}{construct\_pseudobands}
 
\caption{Stochastic Pseudobands}\label{alg:main}

\Input{\float \en[\nk, \nb]  \Comment*[r]{mean-field energies (kpoints, bands)}\\
\nonl\complex \wfn[\nk, \nb, \nbasis] \Comment*[r]{mean-field wavefunctions}\\
\nonl\int \npv, \npc $\geq 0$ \Comment*[r]{protected states (valence/conduction)} \\
\nonl\int \nxv, \nxc $>0$ \Comment*[r]{stochastic states per subspace}\\
\nonl\float $2 \gtrsim$ \fv, \fc $> 0$ \Comment*[r]{constant energy fraction}}

\BlankLine
\Output{\float \enout[\nk, \nspb]  \Comment*[r]{output energies  (kpoints, pseudobands)}\\
\nonl\complex \wfnout[\nk, \nspb, \nbasis] \Comment*[r]{output wavefunctions}}

\BlankLine
\nonl\Begin{
set Fermi level to zero\;

partition \en into valence and conduction states \env, \enc \;
\BlankLine

\Comment{construct slices}

\slicec = \constructslices{\enc, \npc, \fc}\; \Comment*[r]{See Alg.~\ref{alg:slice}}

\slicev = \constructslices{-\env, \npv, \fv}\;

\BlankLine

\Comment{initialize outputs}

initialize output wavefunction \wfnout and energies \enout\;

copy protected states and energies to \wfnout and \enout\;

\BlankLine

\Comment{construct pseudobands}
\constructspb{\wfnout, \enout, \slicec, \enc, \wfn}\; \Comment*[r]{See Alg.~\ref{alg:spb}}

\constructspb{\wfnout, \enout, \slicev, \env, \wfn}\;

\Return \wfnout, \enout
}

\end{algorithm}

\begin{algorithm}
\setcounter{AlgoLine}{0}
\SetAlgoLined
\DontPrintSemicolon
\SetKwData{wfn}{wfn}\SetKwData{wfnout}{wfn\_out}\SetKwData{nk}{n\_k}\SetKwData{nb}{n\_b}\SetKwData{nspb}{n\_spb}\SetKwData{nbasis}{n\_basis}\SetKwData{en}{en}\SetKwData{enout}{en\_out}\SetKwData{env}{en\_v}\SetKwData{envtmp}{en\_v\_tmp}\SetKwData{enc}{en\_c}\SetKwData{first}{first}\SetKwData{last}{last}
\SetKwData{firstidx}{first\_idx}\SetKwData{lastidx}{last\_idx}
\SetKwData{slicec}{slices\_c}\SetKwData{slices}{slices}\SetKwData{slice}{slice}\SetKwData{slicev}{slices\_v}
\SetKwData{nsc}{ns\_c}\SetKwData{np}{np}\SetKwData{npc}{np\_c}\SetKwData{nxc}{n$\xi$\_c}\SetKwData{fc}{$\f$\_c}
\SetKwData{nsv}{ns\_v}\SetKwData{npv}{np\_v}\SetKwData{nxv}{n$\xi$\_v}\SetKwData{fv}{$\f$\_v}\SetKwData{ff}{$\f$}
\SetKwData{complex}{\textbf{complex}}\SetKwData{float}{\textbf{float}}\SetKwData{int}{\textbf{int}}
\SetKwData{ik}{ik}
\SetKwData{shift}{shift}
\SetKwData{sliceidx}{slice\_idx}
\SetKwData{ix}{i$\xi$}
\SetKwData{phases}{phases}
\SetKwData{nbs}{n\_b\_slice}
\SetKwInOut{Input}{input}\SetKwInOut{Output}{output}
\SetKwFunction{min}{min}
\SetKwFunction{len}{len}
\SetKwFunction{mysqrt}{sqrt}
\SetKwFunction{mean}{mean}
\SetKwFunction{range}{range}
\SetKwFunction{matvec}{matvec}
\SetKwFunction{findbandidx}{find\_band\_idx}
\SetKwFunction{append}{append}
\caption{Construction of Slices}\label{alg:slice}

\Input{\en, \np, \ff}

\BlankLine
\Output{\slices}

\BlankLine

\nonl\Begin{

initialize array \slices and \first = \en[\np+1]\;

\While{\first $<$ \min{\en$[:, -1]$}}{
  \last = \first *  (1 + \ff)\;
  
  \firstidx = \findbandidx{\first}\; \Comment*[r]{find index of band with this energy}
  
  \lastidx = \findbandidx{\last}\;
  
  \append{\slices, $[$\firstidx, \lastidx$]$}\;
  
  \first = \en[\lastidx + 1]\;
}
\Return \slices

}
\end{algorithm}

\begin{algorithm}
\setcounter{AlgoLine}{0}
\SetAlgoLined
\DontPrintSemicolon
\SetKwData{wfn}{wfn}\SetKwData{wfnout}{wfn\_out}\SetKwData{nk}{n\_k}\SetKwData{nb}{n\_b}\SetKwData{nspb}{n\_spb}\SetKwData{nbasis}{n\_basis}\SetKwData{en}{en}\SetKwData{enout}{en\_out}\SetKwData{env}{en\_v}\SetKwData{envtmp}{en\_v\_tmp}\SetKwData{enc}{en\_c}\SetKwData{first}{first}\SetKwData{last}{last}
\SetKwData{firstidx}{first\_idx}\SetKwData{lastidx}{last\_idx}
\SetKwData{slicec}{slices\_c}\SetKwData{slices}{slices}\SetKwData{slice}{slice}\SetKwData{slicev}{slices\_v}
\SetKwData{nsc}{ns\_c}\SetKwData{np}{np}\SetKwData{npc}{np\_c}\SetKwData{nxc}{n$\xi$\_c}\SetKwData{fc}{$\f$\_c}
\SetKwData{nsv}{ns\_v}\SetKwData{npv}{np\_v}\SetKwData{nxv}{n$\xi$\_v}\SetKwData{fv}{$\f$\_v}\SetKwData{ff}{$\f$}
\SetKwData{complex}{\textbf{complex}}\SetKwData{float}{\textbf{float}}\SetKwData{int}{\textbf{int}}
\SetKwData{ik}{ik}
\SetKwData{shift}{shift}
\SetKwData{sliceidx}{slice\_idx}
\SetKwData{ix}{i$\xi$}
\SetKwData{phases}{phases}
\SetKwData{nbs}{n\_b\_slice}
\SetKwInOut{Input}{input}\SetKwInOut{Output}{output}
\SetKwFunction{min}{min}
\SetKwFunction{len}{len}
\SetKwFunction{mysqrt}{sqrt}
\SetKwFunction{mean}{mean}
\SetKwFunction{range}{range}
\SetKwFunction{matvec}{matvec}
\SetKwFunction{findbandidx}{find\_band\_idx}
\SetKwFunction{append}{append}
\caption{Construction of Pseudobands}\label{alg:spb}

\Input{\slices, \en, \wfn}

\Output{\wfnout, \enout}

\BlankLine
\nonl\Begin{

\For{\slice $\in$ \slices}{
    set \nbs = \slice[1] - \slice[0]\;
    
    compute index of next pseudoband \shift
    
    \For{\ik $\in$ \range{\nk}}{
        \For{\ix $\in$ \range{\nxc}}{
            generate array of random phases \phases[\nbs]\;

            \phases \ /= \mysqrt{\nxc}\; \Comment*[r]{normalization}
            \BlankLine

            \Comment{matvec to obtain pseudobands}
            \wfnout[\ik, \shift, :] = \matvec{\wfn$[$\ik, \slice, $:]$, \phases}\;
            \BlankLine

            \Comment{output energy is mean energy of \slice}
            \enout[\ik, \shift] = \mean{\en$[$\ik, \slice$]$}
        }
    }
}
\Return{\wfnout, \enout}
}
\end{algorithm}

\newpage
\section{Convergence Tests for Additional Systems}\label{extraconvtests}

\subsection{ZnO Supercells}
\begin{figure}[H]
    \centering
    \includegraphics[width=\textwidth-2in]{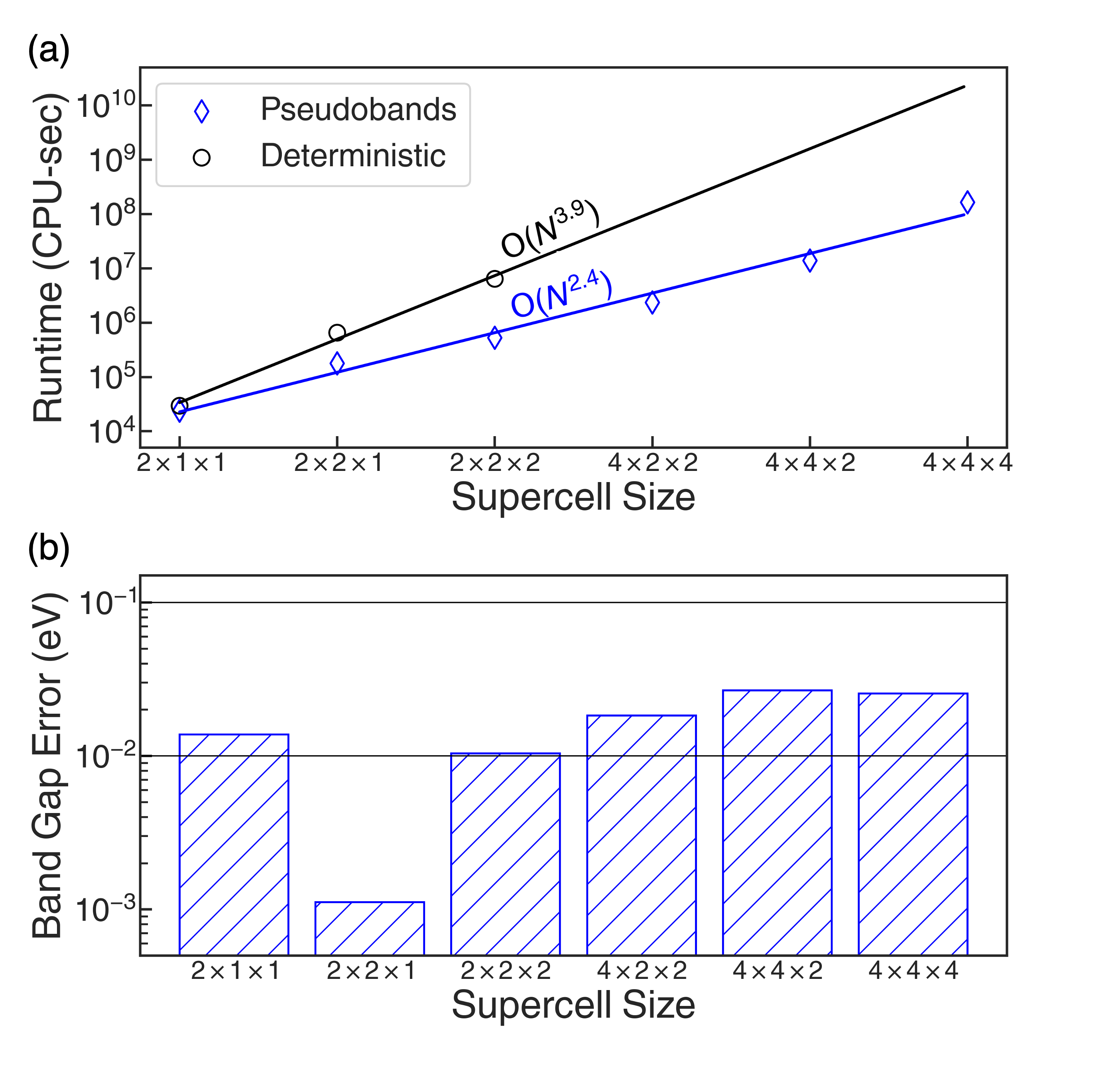}
    \caption{(a) Scaling curve for the dielectric computation per q/k-point for ZnO supercells showing quasi-quadratic behavior up to 256 atoms. (b) QP band gap errors for each calculation with pseudobands; errors are maintained at $<50$ meV for the same convergence parameters $N_\xi, \f$.}
    \label{ZnO_supererror}
\end{figure}

Here we show the band gap error for the calculations of bulk wurtzite ZnO supercells discussed in the main text. In Figure~\ref{ZnO_supererror}(a) below, we reproduce the scaling curve from the main text for ZnO supercells, and in Figure~\ref{ZnO_supererror}(b) we show the band gap error from each supercell calculation resulting from the use of stochastic pseudobands, in reference to the highly converged deterministic unit cell calculation involving 10,000 Kohn-Sham states. As can be seen, except for a serendipitously low error for the $2\times 2\times 1$ supercell, the error is maintained at a constant below 50 meV, without improving the pseudobands parameters $N_\xi$ and $\f$.

\subsection{MoS${}_2$ Bilayer Unit Cell}
\begin{figure}[H]
    \centering
    \includegraphics[width=\textwidth-2in]{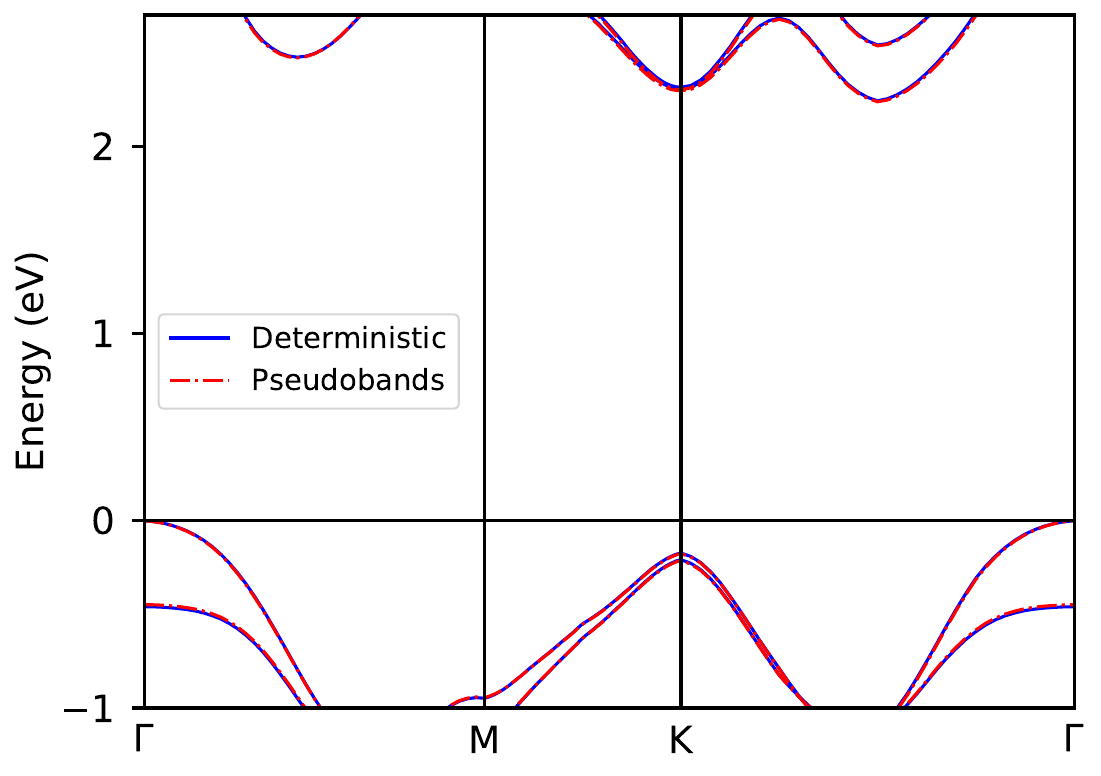}
    \caption{QP band structure of the AA-stacked MoS${}_2$ bilayer. The agreement between the calculation employing stochastic pseudobands and the deterministic approach is nearly perfect, with an RMS error of 20~meV.}
    \label{mos2bands}
\end{figure}

To test convergence in 2D materials, we calculate the band structure of an AA-stacked MoS${}_2$ bilayer. This also serves as preparation for the calculation on a moir\'e superlattice of bilayer MoS$_2$, as described above. Thorough parameter testing was not performed as with the systems in the main text; instead, we chose reasonable parameters based on those systems and performed a single pseudobands calculation, along with the full deterministic calculation (see section S\ref{MoS2details}). The results are shown in Figure \ref{mos2bands} --- with a 20 meV RMS error in the QP energy levels, the two calculations are nearly indistinguishable by eye. Moreover, the pseudobands GW calculation was about 10 times faster, even for this small system.

\subsubsection{Extrapolation of Quasiparticle Corrections for Moiré Systems}

We briefly address a common approach taken to evaluate quasiparticle corrections of moiré bilayers~\cite{lu2019modulated} before the explicit GW calculations allowed by pseudobands approach. For moderate twist angles (so as to not form strong structural inhomogeneity), one expects that the polarizability will not change significantly from that of the untwisted bilayer. For these systems, one may approximate the GW QP energies $E_{\text{moiré}}^{\text{GW}}$ evaluated on the moiré cell as $E_{\text{moiré}}^{\text{GW}} \approx E_{\text{moiré}}^{\text{DFT}} + \delta E_{\text{prim}}^{\text{GW-DFT}}$, where $E_{\text{moiré}}^{\text{DFT}}$ are the DFT Kohn-Sham eigenvalues obtained on the moiré cell, and $E_{\text{prim}}^{\text{GW-DFT}}$ is the GW self-energy correction obtained on the high-symmetry primitive unit cell, relative to the DFT calculation on the equivalent structure.

For the GW calculation of the 5.75$^\circ$ twisted MoS$_2$ bilayer presented in the main text, we find the structurally-driven moiré effects and electron correlation-driven quasiparticle effects are additive to within 100 meV; \textit{i.e.} we find that the difference between the additive approach and explicit calculation of $E_{\text{moiré}}^{\text{GW}}$ varies from 10 meV to 94 meV for selected quasiparticle energies between relevant high-symmetry states close to the Fermi energy. In particular, the splitting $E_{\text{VBM}}(\Gamma) - E_{\text{VBM-1}}(\Gamma)$ is accurately captured by the additive approach, with an error of only 10 meV. However, for the splittings $E_{\text{VBM}}(\Gamma) - E_{\text{VBM}}(K)$ and $E_{\text{CBM}}(K) - E_{\text{CBM}}(\Lambda)$, the error is 94 meV and 42 meV, respectively (see Fig. 3 in the main text), showing that the additive approach cannot be applied everywhere in the moiré cell with uniform accuracy.  Furthermore, one expects that, for other large-scale systems wherein the dielectric function is not uniform within the cell, such as in moiré systems displaying varying degree of hybridization between valence and conduction bands (e.g., bilayer graphene) or when there is charge doping localized within the moiré cell, such an additive procedure may show further inaccuracies.

\subsection{Ag${}_{54}$Pd Nanoparticle}\label{AgNP}
\begin{figure}[H]
    \centering
    \includegraphics[height=\textheight-1.85in]{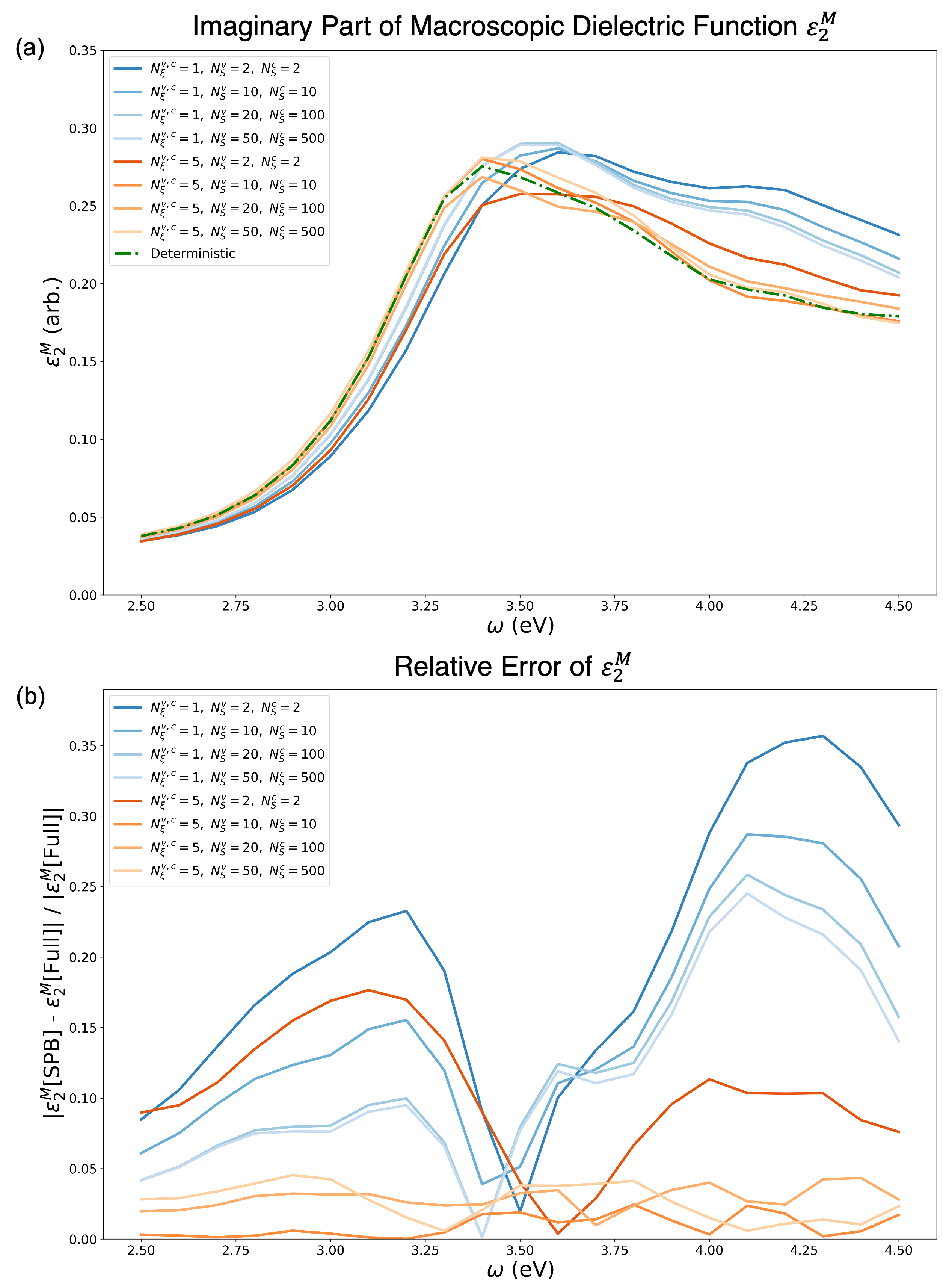}
    \caption{(a) Pseudobands convergence of the imaginary part of the macroscopic dielectric function for an Ag${}_{54}$Pd nanoparticle. All calculations used $N_P^{v/c} = 10$ and other parameters are listed in the legend. Speedups of 10-100 times are observed relative to the deterministic calculation. (b) The same data as (a) but as a percent difference. $\epsilon_2^M[\mathrm{Full}]$ denotes the deterministic calculation.}
    \label{AgNPfig}
\end{figure}

Here we test the convergence of the imaginary part of the macroscopic dielectric function of an Ag${}_{54}$Pd nanoparticle in the full-frequency scheme (see section S\ref{fullfreq}) with respect to $N_S$ and $N_\xi$. In Figure~\ref{AgNPfig}a we show several calculations of the macroscopic dielectric function with varying pseudobands parameters listed in the legend, along with the reference calculation. Figure~\ref{AgNPfig}b shows the same data as Figure~\ref{AgNPfig}a but as a percent error, demonstrating convergence over the whole frequency grid. In Table~\ref{AgNPtable} we summarize the results from Figure~\ref{AgNPfig}, showing the pseudobands parameters, corresponding speedups, and root mean square percentage error (RMSPE). RMSPE is defined as 
\begin{align}
    \mathrm{RMSPE} &= \sqrt{\frac{1}{N_\omega}\sum_{\omega} \left(\ddfrac{\epsilon^M_2(\omega)[\mathrm{SPB}] - \epsilon^M_2(\omega)}{\epsilon^M_2(\omega)}\right)^2}\times 100\%, \label{rmspe}
\end{align}
where for this calculation the sum over $\omega$ ranges from $\omega = 2.5$ eV to $\omega = 4.5$ eV, with $N_\omega = 21$. Untabulated parameters are the same for all runs, and are written in section S\ref{AgNpdetails}. As can be seen, smooth convergence is achieved as the pseudobands parameters become better, demonstrating the ability to handle full-frequency calculations. Further convergence can be achieved by reducing $\delta\omega$, which was 100 meV for these calculations. We emphasize that while the $N_\xi=1$ calculations do converge, this is primarily due to the large values of $N_S$ used and the averaging of the matrix elements discussed in section S\ref{chiderivations} below; the $N_\xi=5$ calculations which actually make use of stochastic averages converge much faster. We also note that in general, the full-frequency dielectric calculation is not as rapidly converging as the dielectric calculation with the GPP model, or the self-energy calculation, so a higher value of $N_\xi$ is recommended. See section S\ref{fullfreq} for details.

\begin{table}[H]
\centering
\begin{tabular}{|c|c|c|c|c|}
 \hline
 \rule{0pt}{4ex} $N_\xi^{v,c}$ & $N_S^v$ & $N_S^c$ & Speedup Factor & RMSPE (\%)\\  [1.5 ex] 
 \hline\hline 
\rule{0pt}{3ex} 1 & 2 & 2 & 84 & 22.2 \\ [1 ex] 
 \hline 
\rule{0pt}{3ex} 1 & 10 & 10 & 86 & 17.2 \\ [1 ex] 
 \hline 
\rule{0pt}{3ex} 1 & 20 & 100 & 80 & 14.6 \\ [1 ex] 
\hline 
\rule{0pt}{3ex} 1 & 50 & 500 & 80 & 13.5 \\ [1 ex] 
\hline
\rule{0pt}{3ex} 5 & 2 & 2 & 65 & 11.2 \\ [1 ex] 
\hline 
\rule{0pt}{3ex} 5 & 10 & 10 & 65 & 1.2 \\ [1 ex] 
\hline 
\rule{0pt}{3ex} 5 & 20 & 100 & 52 & 3.0 \\ [1 ex] 
 \hline 
\rule{0pt}{3ex} 5 & 50 & 500 & 26 & 2.9 \\ [1 ex] 
 \hline 
\end{tabular}
\caption{(rightmost column) Root mean square percent error (RMSPE) of the frequency-dependent imaginary part of the macroscopic dielectric function $\epsilon_2^M(\omega)$ for an Ag${}_{54}$Pd nanoparticle, showing convergence over $N_S$ and $N_\xi$. Other columns show pseudobands parameters used for each calculation and the corresponding speedup factor observed, excluding I/O. Error and speedups are measured against a highly converged deterministic calculation. Small variations are expected in the error and speedups due to different random coefficients $\alpha$ and different node configurations, respectively. Corresponding $\f$ values to the $N_S$ in the table are $\f^v = \{0.51, 
0.11, 0.11, 0.0079\}$ and $\f^c = \{0.67, 0.14, 0.014, 0.0027\}$.}
\label{AgNPtable}
\end{table}

\newpage
\section{Convergence Proofs for GW Quantities with Pseudobands}
In this section, we rigorously prove the intuitive convergence of our method described in the main text. We do so by showing explicitly that the error introduced to the electronic Green's function $G(\omega)$ and the static non-interacting polarizability $\chi^0_{\vb{G,G'}}(\vb{q}, \omega\approx 0)$ goes to 0 upon the usage of stochastic pseudobands. Additionally, we qualitatively discuss the convergence of the self-energy $\Sigma^{GW}$ when we utilize stochastic pseudobands.

We first recall the equations defining pseudobands from the main text:
\begin{align}
    G(\omega) &\equiv \sum_{n,{\vb{k}}} \frac{\ket{\phi_{n{\vb{k}}}}\bra{\phi_{n{\vb{k}}}}}{\omega - E_{n{\vb{k}}} \mp i\eta} \equiv \sum_{{\vb{k}}} G_{{\vb{k}}}(\omega)\label{green}\\
    G_{{\vb{k}}}(\omega) &\approx G_{{\vb{k}}}^P(\omega) + \sum_S^{N_S} G_{{\vb{k}}}^S(\omega)\label{S-Dpart}\\
    G_{{\vb{k}}}^P(\omega) &= \sum_{n \in P}^{N_P} \frac{\ket{\phi_{n\vb{k}}}\bra{\phi_{n\vb{k}}}}{\omega -E_{n\vb{k}} \mp i\eta}\\
    G_{{\vb{k}}}^S(\omega) &= \frac{1}{\omega - \bar E_{S\vb{k}} \mp i\eta}\sum_{i=1}^{N_\xi}\ket{\xi^S_{i,{\vb{k}}}}\bra{\xi^S_{i,{\vb{k}}}} \label{SPBappx}\\
    \ket{\xi^S_{i, \vb{k}}} &= \frac{1}{\sqrt{N_\xi}}\sum_{n\in S}\alpha^S_{i, n\vb{k}}\ket{\phi_{n\vb{k}}}\label{SPBdef}.
\end{align}

Here $\ket{\phi_{n\vb{k}}}$ are Kohn-Sham states, $ E_{n{\vb{k}}}$ are the corresponding mean-field eigenenergies, $\eta = 0^+$, $P$ is the protected subspace, $S$ is a stochastic subspace of which there are $N_S$, $\bar E_{S\vb{k}}$ is the average energy of the Kohn-Sham states in subspace $S$, $\ket{\xi^S_{i,{\vb{k}}}}$ is a stochastic vector in subspace $S$ of which we take $N_\xi$ to resolve the projection onto subspace $S$, and $\alpha^S_{i, n\vb{k}}$ are uniformly distributed random phases with which we construct the pseudobands $\ket{\xi^S_{i,{\vb{k}}}}$. The stochastic subspaces $S$ can run over unoccupied and occupied states, and we treat the general case here. We do not make assumptions about the distribution of subspaces $S$ in terms of how many states each subspace holds at this point. Rather, we will derive a partition that allows rapid convergence of the quantities of interest (see section~\ref{partition}).

In the following subsections, we show convergence of $G$ and $\chi^0(\omega\approx 0)$, discuss convergence of $\Sigma^{GW}$, derive a heuristic for partitioning subspaces, and extend the approach to finite $\omega\neq 0$.

\subsection{Convergence of the Green's Function $G(\omega)$} \label{greenconvderivation}

Below, we denote the Green's function evaluated with stochastic pseudobands (SPB) as $G(\omega)[\mathrm{SPB}]$. The convergence of $G(\omega)[\mathrm{SPB}]$ is not smooth in the sense that, if we define the error of stochastic pseudobands for $G$ as 
\begin{align}
    \mathrm{Err}\left[G(\omega)[\mathrm{SPB}]\right] &\equiv G(\omega)[\mathrm{SPB}] - G(\omega)
\end{align}
and take the limit $\eta \rightarrow 0$, we always have poles in $\mathrm{Err}\left[G(\omega)[\mathrm{SPB}]\right]$ at every energy $E_{n\vb{k}}$ of the original mean-field states. This is because we approximate a number of densely-packed poles of $G$ with a single pole of their combined weight in a given subspace $S$, we always have poles in our error coming from the true $G$.

However, if we assume $\omega$ is not close to a pole of either $G(\omega)[\mathrm{SPB}]$ or $G(\omega)$, then we can obtain expressions for the expectation and variance of $\mathrm{Err}\left[G(\omega)[\mathrm{SPB}]\right]$ and show that both go to 0.

First, for a given subspace $S$, we define $G[S]$ as the restriction of $G[\mathrm{SPB}]$ to $S$, and find
\begin{align}
    \mathbb{E}\left[\mathrm{Err}\left[G_{\vb{k}}(\omega)[S]\right]\right] &= \sum_{n\in S} \ket{\phi_{n\vb{k}}}\bra{\phi_{n\vb{k}}}\left[\frac{1}{\omega - \bar E_{S,\vb{k}} \mp i\eta} - \frac{1}{\omega - E_{n\vb{k}} \mp i\eta}\right],
\end{align}
since the stochastic off-diagonals have an expectation of 0. Thus, the main effect of the constant energy denominator approximation is to slightly change the expected value of the Green's function coming from subspace $S$. Since $\bar{E}_{S,\vb{k}}-E_{n,\vb{k}}\lesssim \Delta E_S$, the energy range spanned by $S$, this error is roughly proportional to $\Delta E_S / (\omega - \bar{E}_{S,\vb{k}})^2$, and decreases as the number of slices increases, $N_S\rightarrow \infty$. 

Next, we move on to the variance of the error of $G[S]$, to which only the off-diagonal terms in the Kohn-Sham basis contribute. We obtain, for a given matrix element $n\neq n'\in S$ for a given subspace $S$,
\begin{align}
    \var\left[\mathrm{Err}\left[G_{n,n',\vb{k}}(\omega)[S]\right]\right] &= \frac{1}{\left(\omega - \bar E_{S,\vb{k}}\right)^2}\mathbb{E}\left[\left|\frac{1}{N_\xi}\sum_{i=1}^{N_\xi}\alpha^S_{n,i}\left(\alpha^S_{n',i}\right)^*\right|^2\right]\\
    &= \frac{1}{\left(\omega - \bar E_{S,\vb{k}}\right)^2} \cdot \frac{1}{N_\xi},
\end{align}
which approaches $0$ with increasing stochastic pseudobands, $N_\xi \rightarrow \infty$. Thus, with both the expectation and variance of the error converging to 0 as $N_S\rightarrow \infty$ and $N_\xi \rightarrow \infty$, we have convergence of our proposed approach when evaluating the electronic Green's function away from a pole in any subspace.

In practice however, we take $\eta > 0$ to be a small finite number on the order of 100 meV which broadens and smooths out the poles by shifting them away from the real axis, so convergence can be achieved at any $\omega$. This is the standard practice, for instance, when evaluating the frequency-dependent dielectric function. With this finite broadening, the convergence of $G(\omega)$ is much smoother than in the formal $\eta\rightarrow 0$ limit, and the usage of stochastic pseudobands is well behaved at any frequency. 

Additionally, we see empirically that as few as $N_\xi = 2$ or $3$ pseudobands are typically enough to get $< 100$~meV error in the QP energies, which is a surprisingly low parameter given the $1/N_\xi$ convergence of the Green's function itself. As discussed in the following section, the polarizability $\chi^0(\omega\approx 0)[\mathrm{SPB}]$ tends to converge much faster than $G[\mathrm{SPB}]$, partially due to the rapidly oscillating nature of the matrix elements involving Kohn-Sham states used in the evaluation of the polarizability.

\subsection{Convergence of the Non-interacting Polarizability $\chi^0$}\label{chiderivations}

Here we compute, in detail, the error of $\chi^0(\omega\approx0)[\mathrm{SPB}]$, the static non-interacting polarizability evaluated with stochastic pseudobands. While the following is done in a plane-wave basis, a similar analysis should hold in other bases. We start from the Adler-Wiser formula for the non-interacting polarizability matrix~\cite{hybertsen1986electron},
\begin{align}
\chi^0_{\G,\G'}(\vb{q}, \omega=0) &= \sum_{vc\vb{k}} \ddfrac{\bra{\phi_{c\vb{k}}} e^{-i(\vb{q}+\G)\cdot \vb{r}}\ket{\phi_{v\vb{k}+\vb{q}}} \bra{\phi_{v\vb{k}+\vb{q}}} e^{i(\vb{q}+\G')\cdot \vb{r}}\ket{\phi_{c\vb{k}}}}{E_{v\vb{k}+\vb{q}} - E_{c\vb{k}}} \equiv \sum_{\vb{k}}\chi^0_{\G,\G',\vb{k}}(\vb{q}, 0), \label{adler-wiser}
\end{align}
where we focus for now on $\omega = 0$ as used in the GPP, where $\vb{G,G'}$ are reciprocal lattice vectors, $v$ and $c$ denote valence and conduction bands, respectively, and where we have again partitioned over $\vb{k}$-points. For convenience, we define $M_{vc\vb{k}}(\vb{q},\G') = \bra{\phi_{v\vb{k}+\vb{q}}} e^{i(\vb{q}+\G')\cdot \vb{r}}\ket{\phi_{c\vb{k}}}$.

With the usage of stochastic pseudobands, we can partition Eq.~(\ref{adler-wiser}) into four components depending on whether $v$ or $c$ is deterministic or stochastic,
\begin{align}
\begin{split}
    \chi^0_{\G,\G',\vb{k}}(\vb{q}, 0)[\mathrm{SPB}] &= \chi^0_{\G,\G'}(\vb{q}, 0)[v \leq N_P^v; c \leq N_P^c]\\ & + \chi^0_{\G,\G'}(\vb{q}, 0)[v \leq N_P^v; c\in S^c] + \chi^0_{\G,\G'}(\vb{q}, 0)[v \in S^v; c\leq N_P^c]\\
    & + \chi^0_{\G,\G'}(\vb{q}, 0)[v \in S^v; c\in S^c],
    \label{chi_split}
\end{split}
\end{align}
where we take the indexing convention that we count up from the Fermi level for both valence and conduction states. Then, we can define the error of the polarizability with pseudobands as
\begin{align}
    \mathrm{Err}\Big[\chi^0_{\G,\G',\vb{k}}(\vb{q}, 0)[\mathrm{SPB}]\Big] &\equiv \chi^0_{\G,\G',\vb{k}}(\vb{q}, 0)[\mathrm{SPB}] - \chi^0_{\G,\G',\vb{k}}(\vb{q}, 0).\label{chierr}
\end{align}
As for $G$, to prove convergence we compute the expectation and variance of the error Eq.~(\ref{chierr}) and show both tend to 0 as $N_S, N_\xi \rightarrow \infty$. 

Clearly, the first term in Eq.~(\ref{chi_split}) contributes no error. So, we focus on the last three terms, starting with $\chi^0_{\G,\G'}(\vb{q}, 0)[v \leq N_P^v; c\in S^c]$. Explicitly, we have for the matrix elements with pseudobands:
\begin{align}
    \bra{\phi_{v\vb{k}+\vb{q}}} e^{i(\vb{q}+\G')\cdot \vb{r}}\ket{\xi^{S^c}_{i\vb{k}}} &= \frac{1}{\sqrt{N_\xi}}\sum_{n\in S^c} \alpha_{i,n\vb{k}}^{S^c}\bra{\phi_{v\vb{k}+\vb{q}}} e^{i(\vb{q}+\G')\cdot \vb{r}}\ket{\phi_{n\vb{k}}}.
\end{align}
Restricting to a single valence band $v$ and k-point $\vb{k}$, we have for the error:
\begin{align}
\begin{split}
    \mathrm{Err}\Big[\chi^0_{\G,\G', \vb{k}}(\vb{q}, 0)[v; c\in S^c]\Big] &= \frac{1}{N_\xi}\sum_{i=1}^{N_\xi}\sum_{n, n' \in S^c} \left(\alpha_{i,n\vb{k}}^{S^c}\right)^* \alpha_{i,n'\vb{k}}^{S^c}\bra{\phi_{n\vb{k}}} e^{-i(\vb{q}+\G)\cdot \vb{r}}\ket{\phi_{v\vb{k}+\vb{q}}} \\ & \hspace{20pt} \times \bra{\phi_{v\vb{k}+\vb{q}}} e^{i(\vb{q}+\G')\cdot \vb{r}}\ket{\phi_{n'\vb{k}}} \times \ddfrac{1}{E_{v\vb{k}+\vb{q}} - \bar E_{S^c\vb{k}}} \\
    & - \sum_{n \in S^c} \bra{\phi_{n\vb{k}}} e^{-i(\vb{q}+\G)\cdot \vb{r}}\ket{\phi_{v\vb{k}+\vb{q}}} \\ & \hspace{20pt} \times \bra{\phi_{v\vb{k}+\vb{q}}} e^{i(\vb{q}+\G')\cdot \vb{r}}\ket{\phi_{n\vb{k}}} \times \ddfrac{1}{E_{v\vb{k}+\vb{q}} - E_{n\vb{k}}} .
\end{split}\label{err_chi_v_Sc}
\end{align}
We can partition this into diagonal and off-diagonal terms so that we can treat the mean-energy error and stochastic error independently:
\begin{align}
    \mathrm{Err}\Big[\chi^0_{\G,\G', \vb{k}}(\vb{q}, 0)[v; c\in S^c]\Big] &= \left[ n = n'\right] +  \left[ n \neq n'\right] . \label{triangle}
\end{align}
The first term of Eq.~(\ref{triangle}) has no stochastic coefficients, and corresponds only to the mean-energy error:
\begin{align}
\begin{split}
   \left[ n = n'\right] &=  \sum_{n \in S^c} \bra{\phi_{n\vb{k}}} e^{-i(\vb{q}+\G)\cdot \vb{r}}\ket{\phi_{v\vb{k}+\vb{q}}} \bra{\phi_{v\vb{k}+\vb{q}}} e^{i(\vb{q}+\G')\cdot \vb{r}}\ket{\phi_{n\vb{k}}} \\& \hspace{50pt} \times \left[\ddfrac{1}{E_{v\vb{k}+\vb{q}} - \bar E_{S^c\vb{k}}} - \ddfrac{1}{E_{v\vb{k}+\vb{q}} - E_{n\vb{k}}}\right].
\end{split}
\end{align}
We define $\delta E_{n,\vb{k}} \equiv E_{n\vb{k}} - \bar E_{S^c\vb{k}}$ to write:
\begin{align}
   \left[ n = n'\right]  &=  \sum_{n \in S^c} M_{vn\vb{k}}(\vb{q},\G)^*M_{vn\vb{k}}(\vb{q},\G')\times \left[\ddfrac{1}{E_{v\vb{k}+\vb{q}} - \bar E_{S^c\vb{k}}} - \ddfrac{1}{E_{v\vb{k}+\vb{q}} - \delta E_{n,\vb{k}} - \bar E_{S^c\vb{k}}}\right]\\
\begin{split}
    &= \sum_{n \in S^c} M_{vn\vb{k}}(\vb{q},\G)^*M_{vn\vb{k}}(\vb{q},\G')\\
    & \hspace{30pt}\times \left[\ddfrac{1}{E_{v\vb{k}+\vb{q}} - \bar E_{S^c\vb{k}}} - \ddfrac{1}{E_{v\vb{k}+\vb{q}} - \bar E_{S^c\vb{k}}}\cdot \ddfrac{1}{1 - \ddfrac{\delta E_{n,\vb{k}}}{E_{v\vb{k}+\vb{q}} - \bar E_{S^c\vb{k}}}}\right]
\end{split}\\[20pt]
\begin{split}
    &\approx \ddfrac{-1}{\left(E_{v\vb{k}+\vb{q}} - \bar E_{S^c\vb{k}}\right)^2}\sum_{n \in S^c} M_{vn\vb{k}}(\vb{q},\G)^*M_{vn\vb{k}}(\vb{q},\G') \times \delta E_{n,\vb{k}},\label{expectation_v_Sc}
\end{split}
\end{align}
where in the last line we have taken a first-order expansion around $\delta E_{n,\vb{k}} = 0$. This expansion is always valid as the energy denominator $|E_{v\vb{k}+\vb{q}} - \bar E_{S^c\vb{k}}| > 0$. This $\left[ n = n'\right] $ term contributes only to the expectation of the error, and vanishes as $N_S \rightarrow \infty$.

Now we turn to the off-diagonal contribution $\left[ n \neq n'\right] $ of Eq.~(\ref{triangle}):
\begin{align}
\begin{split}
    \left[ n \neq n'\right] &= \frac{1}{N_\xi}\sum_{i=1}^{N_\xi}\sum_{n\neq n' \in S^c} \left(\alpha_{i,n\vb{k}}^{S^c}\right)^* \alpha_{i,n'\vb{k}}^{S^c}\bra{\phi_{n\vb{k}}} e^{-i(\vb{q}+\G)\cdot \vb{r}}\ket{\phi_{v\vb{k}+\vb{q}}} \\ & \hspace{20pt} \times \bra{\phi_{v\vb{k}+\vb{q}}} e^{i(\vb{q}+\G)\cdot \vb{r}}\ket{\phi_{n'\vb{k}}} \times \ddfrac{1}{E_{v\vb{k}+\vb{q}} - \bar E_{S^c\vb{k}}}.
\end{split}
\end{align}
It is easy to show that the distribution of the random variable $\left(\alpha_{i,n\vb{k}}^{S^c}\right)^* \alpha_{i,n'\vb{k}}^{S^c}$ is uniform, the same as its factors. Thus, the expectation is simply 0:
\begin{align}
    \mathbb{E} \left[ n \neq n'\right] &= 0.
\end{align}

By analogy, we obtain the following expressions for $\chi^0_{\G,\G'}(\vb{q}, 0)[v \in S^v; c]$:
\begin{align}
    \left[ n = n'\right] &\approx \ddfrac{1}{\left(\bar E_{S^v\vb{k}+\vb{q}} - E_{c\vb{k}}\right)^2}\sum_{n \in S^v} M_{nc\vb{k}}(\vb{q},\G)^*M_{nc\vb{k}}(\vb{q},\G') \times \delta E_{n,\vb{k}+\vb{q}} \label{expectation_Sv_c}
\end{align}
and 
\begin{align}
    \mathbb{E} \left[ n \neq n'\right] &= 0.
\end{align}

For the fully stochastic term $\chi^0_{\G,\G'}(\vb{q}, 0)[v \in S^v; c\in S^c]$ we have
\begin{align}
\begin{split}
    \mathrm{Err}\Big[\chi^0_{\G,\G', \vb{k}}(\vb{q}, 0)[v\in S^v; c\in S^c]\Big] &= \Bigg[\frac{1}{N_\xi^2}\sum_{i,j=1}^{N_\xi}\sum_{\substack{v, v' \in S^v\\c, c' \in S^c}} \alpha_{i,v\vb{k}+\vb{q}}^{S^v} \left(\alpha_{i,v'\vb{k}+\vb{q}}^{S^v}\right)^*\left(\alpha_{j,c\vb{k}}^{S^c}\right)^* \alpha_{j,c'\vb{k}}^{S^c}\\ & \hspace{-20pt} \times \bra{\phi_{c\vb{k}}} e^{-i(\vb{q}+\G)\cdot \vb{r}}\ket{\phi_{v\vb{k}+\vb{q}}} \bra{\phi_{v'\vb{k}+\vb{q}}} e^{i(\vb{q}+\G)\cdot \vb{r}}\ket{\phi_{c'\vb{k}}} \times \ddfrac{1}{\bar E_{S^v\vb{k}+\vb{q}} - \bar E_{S^c\vb{k}}}\Bigg] \\
    & \hspace{-30pt} - \Bigg[ \sum_{\substack{v\in S^v\\c \in S^c}} \bra{\phi_{c\vb{k}}} e^{-i(\vb{q}+\G)\cdot \vb{r}}\ket{\phi_{v\vb{k}+\vb{q}}} \bra{\phi_{v\vb{k}+\vb{q}}} e^{i(\vb{q}+\G)\cdot \vb{r}}\ket{\phi_{c\vb{k}}} \times \ddfrac{1}{E_{v\vb{k}+\vb{q}} - E_{c\vb{k}}} \Bigg].
\end{split}\label{err_chi_Sv_Sc}
\end{align}
Again for any term not on the diagonal we have 
\begin{align}
    \mathbb{E} \left[\textrm{off-diagonal}\right] &= 0
\end{align}
For the diagonal we have
\begin{align}
    \left[v = v'; c = c'\right] &=  \sum_{\substack{v\in S^v\\c \in S^c}} M_{vc\vb{k}}(\vb{q},\G)^*M_{vc\vb{k}}(\vb{q},\G')\times \left[\ddfrac{1}{\bar E_{S^v\vb{k}+\vb{q}} - \bar E_{S^c\vb{k}}} - \ddfrac{1}{\bar E_{S^v\vb{k}+\vb{q}} + \delta E_{v,\vb{k} + \vb{q}} - \bar E_{S^c\vb{k}} - \delta E_{c,\vb{k}}}\right]\\
\begin{split}
    &\approx \ddfrac{1}{\left(\bar E_{S^v\vb{k}+\vb{q}} - \bar E_{S^c\vb{k}}\right)^2}\sum_{\substack{v\in S^v\\c \in S^c}}M_{vc\vb{k}}(\vb{q},\G)^*M_{vc\vb{k}}(\vb{q},\G')\times \left(\delta E_{v,\vb{k} + \vb{q}} - \delta E_{c,\vb{k}}\right).\label{expectation_Sv_Sc}
\end{split}
\end{align}
From expressions~(\ref{expectation_v_Sc}),~(\ref{expectation_Sv_c}), and~(\ref{expectation_Sv_Sc}) we see that convergence of the mean-energy error is achieved as $N_S\rightarrow \infty$. In particular, the sums in these expressions are always bounded from above and below by a value proportional to $\pm\mathcal{M}^2\cdot \Delta E_S \cdot \dim(S)^2$. Here $\mathcal{M}$ is the maximum magnitude of all matrix elements, $\Delta E_S$ is the energy range spanned by $S$, and $\dim(S)$ is the dimension of subspace $S$. As $N_S\rightarrow \infty$, $\Delta E_S\rightarrow 0$ and $\dim(S) \rightarrow 0$, and convergence of the expectation of the polarizability evaluated with pseudobands is achieved. 

Now, we turn to the variance of the error. This is a bit more tedious because, e.g., the products $\alpha_{i,v\vb{k}+\vb{q}}^{S^v} \left(\alpha_{i,v'\vb{k}+\vb{q}}^{S^v}\right)^*\left(\alpha_{j,c\vb{k}}^{S^c}\right)^* \alpha_{j,c'\vb{k}}^{S^c}$ and $\alpha_{i,v\vb{k}+\vb{q}}^{S^v} \left(\alpha_{i,v'\vb{k}+\vb{q}}^{S^v}\right)^*$ are not necessarily independent random variables. So, we calculate
\begin{align}
    \var\left[  \mathrm{Err}\Big[\chi^0_{\G,\G',\vb{k}}(\vb{q}, 0)[\mathrm{SPB}]\Big] \right] &= \mathbb{E}\left[ \left| \mathrm{Err}_{\textrm{off-diagonal}}\Big[\chi^0_{\G,\G',\vb{k}}(\vb{q}, 0)[\mathrm{SPB}]\Big]\right|^2\right], \label{vardef}
\end{align}
where this is the standard definition $\var[Z] = \mathbb{E}\left[|Z - \mathbb{E}[Z]|^2\right]$, and we have subtracted the previously computed expectations to obtain $\mathrm{Err}_{\textrm{off-diagonal}}$, which consists only of off-diagonal pseudobands contributions to $\chi^0_{\G,\G',\vb{k}}(\vb{q}, 0)[\mathrm{SPB}]$.

Explicitly, we have (all terms below are implied to have only off-diagonal contributions over the band indices in their sums):
\begin{subequations}
\begin{align}
     &\hspace{10pt}\var\left[  \mathrm{Err}\Big[\chi^0_{\G,\G',\vb{k}}(\vb{q}, 0)[\mathrm{SPB}]\Big] \right] \nonumber\\&= \mathbb{E}\Big[[v \leq N_P^v; c\in S^c]^2 + [v \in S^v; c\leq N_P^c]^2 + [v \in S^v; c\in S^c]^2\Big]\label{vars}\\
     &+ \mathbb{E}\Big[[v \leq N_P^v; c\in S^c]\left([v \in S^v; c\leq N_P^c]\right)^* + [v \in S^v; c\leq N_P^c]\left([v \leq N_P^v; c\in S^c]\right)^*\Big]\label{indepcovars}\\
     &+ \mathbb{E}\Big[[v \leq N_P^v; c\in S^c]\left([v \in S^v; c\in S^c]\right)^* + [v \in S^v; c\leq N_P^c]\left([v \in S^v; c\in S^c]\right)^*\Big]\label{depcovars1}\\
     &+ \mathbb{E}\Big[[v \in S^v; c\in S^c]\left([v \leq N_P^v; c\in S^c]\right)^* + [v \in S^v; c\in S^c]\left([v \in S^v; c\leq N_P^c]\right)^*\Big]. \label{depcovars2}
\end{align}
\end{subequations}

Above, Eq.~(\ref{vars}) contains the variances of each summand in Eq.~(\ref{chi_split}), Eq.~(\ref{indepcovars}) contains covariances of terms that are independent, and Eqs.~(\ref{depcovars1},~\ref{depcovars2}) contain non-independent covariances. 

First we deal with Eq.~(\ref{vars}) by computing the following covariance:
\begin{align}
\begin{split}
    \cov\left[\left(\left(\alpha_{i,n\vb{k}}^{S}\right)^* \alpha_{i,n'\vb{k}}^{S}\right), \left(\left(\alpha_{i,n''\vb{k}}^{S}\right)^* \alpha_{i,n'''\vb{k}}^{S}\right)\right] &= \mathbb{E}\left[\left(\alpha_{i,n\vb{k}}^{S}\right)^* \alpha_{i,n'\vb{k}}^{S} \alpha_{i,n''\vb{k}}^{S} \left(\alpha_{i,n'''\vb{k}}^{S}\right)^*\right]\\
    &=\begin{cases} 
      1 & n=n''; n'=n''' \\
      0 & \mathrm{otherwise}
   \end{cases}.
\end{split}
\end{align}
Analogously,
\begin{align}
\begin{split}
    &\cov\left[\left(\alpha_{i,v\vb{k}+\vb{q}}^{S^v} \left(\alpha_{i,v'\vb{k}+\vb{q}}^{S^v}\right)^*\left(\alpha_{j,c\vb{k}}^{S^c}\right)^* \alpha_{j,c'\vb{k}}^{S^c}\right), \left(\alpha_{i,v''\vb{k}+\vb{q}}^{S^v} \left(\alpha_{i,v'''\vb{k}+\vb{q}}^{S^v}\right)^*\left(\alpha_{j,c''\vb{k}}^{S^c}\right)^* \alpha_{j,c'''\vb{k}}^{S^c}\right)\right]\\ &= \mathbb{E}\left[\alpha_{i,v\vb{k}+\vb{q}}^{S^v} \left(\alpha_{i,v'\vb{k}+\vb{q}}^{S^v}\right)^*\left(\alpha_{j,c\vb{k}}^{S^c}\right)^* \alpha_{j,c'\vb{k}}^{S^c}\left(\alpha_{i,v''\vb{k}+\vb{q}}^{S^v}\right)^*\alpha_{i,v'''\vb{k}+\vb{q}}^{S^v}\alpha_{j,c''\vb{k}}^{S^c}\left(\alpha_{j,c'''\vb{k}}^{S^c}\right)^* \right]\\
    &=\begin{cases} 
      1 & v=v''; v'=v'''; c=c''; c'=c''' \\
      0 & \mathrm{otherwise}
   \end{cases}
\end{split}
\end{align}
Thus, the variance of the terms in Eq.~(\ref{vars}) can be computed as the sum of variances of their summands over $n$ and $n'$:
\begin{subequations}
\begin{align}
    \mathbb{E}\Big[[v \leq N_P^v; c\in S^c]^2\Big] &= \frac{1}{N_\xi}\sum_{S^c}\sum_{\substack{v \leq N_P^v\\c\neq c'\in S^c}} \left|\bra{\phi_{c\vb{k}}} e^{-i(\vb{q}+\G)\cdot \vb{r}}\ket{\phi_{v\vb{k}+\vb{q}}} \right|^2 \left|\bra{\phi_{v\vb{k}+\vb{q}}} e^{i(\vb{q}+\G)\cdot \vb{r}}\ket{\phi_{c'\vb{k}}}\right|^2\notag\\
    &\times\ddfrac{1}{\left(E_{v\vb{k}+\vb{q}} - \bar E_{S^c\vb{k}}\right)^2}\\
    \mathbb{E}\Big[[v \in S^v; c\leq N_P^c]^2\Big] &= \frac{1}{N_\xi}\sum_{S^v}\sum_{\substack{c \leq N_P^c\\v\neq v'\in S^v}} \left|\bra{\phi_{c\vb{k}}} e^{-i(\vb{q}+\G)\cdot \vb{r}}\ket{\phi_{v\vb{k}+\vb{q}}} \right|^2 \left|\bra{\phi_{v'\vb{k}+\vb{q}}} e^{i(\vb{q}+\G)\cdot \vb{r}}\ket{\phi_{c\vb{k}}}\right|^2\notag\\
    &\times\ddfrac{1}{\left(\bar E_{S^v\vb{k}+\vb{q}} - E_{c\vb{k}}\right)^2}\\
    \mathbb{E}\Big[[v \in S^v; c\in S^c]^2\Big] &= \frac{1}{N_\xi^2}\sum_{S^v, S^c}\sum_{\substack{v\neq v' \in S^v\\ \mathrm{or}\\c\neq c' \in S^c}} \left|\bra{\phi_{c\vb{k}}} e^{-i(\vb{q}+\G)\cdot \vb{r}}\ket{\phi_{v\vb{k}+\vb{q}}} \right|^2\left|\bra{\phi_{v'\vb{k}+\vb{q}}} e^{i(\vb{q}+\G)\cdot \vb{r}}\ket{\phi_{c'\vb{k}}}\right|^2\notag\\
    &\times\ddfrac{1}{\left(\bar E_{S^v\vb{k}+\vb{q}} - \bar E_{S^c\vb{k}}\right)^2}
\end{align}
\label{diagvars}
\end{subequations}

We can see that all these terms go as $1/N_\xi$ or smaller, following the same convergence trend as the stochastic resolution of the identity. Next, we compute the remaining covariances. As noted before, Eq.~(\ref{indepcovars}) contains independent random variables, so these terms are 0 ($\alpha_v$ cannot cancel out $\alpha_c$ and vice versa). For Eqs.~(\ref{depcovars1},~\ref{depcovars2}) we have terms of the following form:
\begin{align}
    \mathbb{E}\left[\left(\alpha_{j,c\vb{k}}^{S^c}\right)^* \alpha_{j,c'\vb{k}}^{S^c}\left(\alpha_{i,v\vb{k}+\vb{q}}^{S^v}\right)^*\alpha_{i,v'\vb{k}+\vb{q}}^{S^v}\alpha_{j,c''\vb{k}}^{S^c}\left(\alpha_{j,c'''\vb{k}}^{S^c}\right)^* \right]&=0,
\end{align}
as there is always a pair of phases ($v,v'$ in this case) which cannot be cancelled out. So, despite the explicit dependence of some of these products, they still have zero covariance. With this, we conclude that the variance of the polarizability is actually given by the sum of Eqs.~(\ref{diagvars}). Thus, we obtain that
\begin{subequations}
\begin{align}
    \var\left[  \mathrm{Err}\Big[\chi^0_{\G,\G',\vb{k}}(\vb{q}, 0)[\mathrm{SPB}]\Big] \right] &\propto \frac{C_1}{N_\xi} + \frac{C_2}{N_\xi^2}\\
    \implies \lim_{N_\xi\rightarrow \infty} \var\left[  \mathrm{Err}\Big[\chi^0_{\G,\G',\vb{k}}(\vb{q}, 0)[\mathrm{SPB}]\Big] \right] &= 0 \label{vartozero}
\end{align}
\end{subequations}
for some constants $C_1, C_2$ that can depend on $N_S$ but are always finite. Thus, Eq.~(\ref{vartozero}) holds for any value of $N_S$. With the expectation of the error also going to 0 as $N_S\rightarrow \infty$ as shown before, \textit{this completes the proof of convergence of the static polarizability with pseudobands.} 

We note that, in practice, due to the rapidly oscillating nature of matrix elements $M_{vc\vb{k}}(\vb{q},\vb{G}')$ with respect to $v$ and $c$, there is a large amount of averaging out in expressions~(\ref{expectation_v_Sc}),~(\ref{expectation_Sv_c}), and~(\ref{expectation_Sv_Sc}). This enables convergence of the expectations derived even with modest values of $N_S$ and $N_\xi$, i.e., the average case is much better than the worst case. 

In section~\ref{fullfreq} we provide a framework for extending the usage of stochastic pseudobands to $\omega \neq 0$, which we verify numerically in section S\ref{AgNP}. Note that the above derivations can be extended to small $\omega$ up to roughly $\min\big(E_{\mathrm{CBM}} - E_{v=N_P^v},\allowdisplaybreaks E_{c=N_P^c} - E_{\mathrm{VBM}}\big)$, as $\omega$ bounded by this value do not produce any divergences in the above expressions.

\subsubsection{Partition of Subspaces}\label{partition}
Here we derive a physically-motivated heuristic for how to partition the mean-field states $\ket{\phi_{n\vb{k}}}$ into the stochastic subspaces $\{S\}$, using the expressions derived in the previous section. While optimizing this partition over the error of $\chi^0$, for example, would be very difficult, it is also not guaranteed that this optimum would minimize the error of $\Sigma^{GW}$ anyways. Instead, we simply demand that each stochastic subspace should contribute a roughly constant error to $\chi^0(\omega=0)$. We saw before that $N_S$ is primarily responsible for controlling the expectation value of $\chi^0[\mathrm{SPB}]$, while $N_\xi$ is responsible for controlling the variance. Therefore, we only focus on the expectation value for this problem. Additionally, we ignore the contribution of Eq.~(\ref{expectation_Sv_Sc}) due to its larger energy denominator. Therefore, we focus on Eq.~(\ref{expectation_v_Sc}) (Eq.~(\ref{expectation_Sv_c}) behaves similarly), and derive the following bounds:
\begin{align}
    \begin{split}
     \left|\mathbb{E}\left[\mathrm{Err}\Big[\chi^0_{\G,\G', \vb{k}}(\vb{q}, 0)[v; c\in S^c]\Big]\right] \right| &\approx \ddfrac{1}{\left(E_{v\vb{k}+\vb{q}} - \bar E_{S^c\vb{k}}\right)^2}\left|\sum_{n \in S^c} M_{vn\vb{k}}(\vb{q},\G)^*M_{vn\vb{k}}(\vb{q},\G') \times \delta E_{n,\vb{k}}\right|\\
     &\leq  \ddfrac{\mathcal{M}^2}{\left(E_{v\vb{k}+\vb{q}} - \bar E_{S^c\vb{k}}\right)^2}\sum_{n \in S^c} \left|\delta E_{n,\vb{k}}\right|\\
     &\leq \ddfrac{\mathcal{M}^2}{\left(\bar E_{S^c\vb{k}}\right)^2}\cdot \Delta E_{S^c\vb{k}} \cdot \dim(S^c), 
\end{split}
\end{align}
where in the last line we used that the Fermi level is our reference energy $E_F = 0$. In 3D, $\dim(S^c) = \int g(E)dE \sim \bar{E}_S^{1/2} \Delta E_S$ where $g$ is the density of states, so we want $\Delta E_S^{2}/\bar{E}_S^{3/2}$ to be a constant to accrue constant error in each slice. To simplify this condition and be dimension-independent, we instead take $\Delta E_S/\bar E_S$ to be a constant as a practical prescription.

With these approximations we have a simple and \textit{exponential} partition of the total energy range into stochastic subspaces defined by the convergence parameter $\f$,
\begin{align}
    \dfrac{\Delta E_S}{\bar E_S} \equiv \f = \mathrm{const}. \label{exppartition}
\end{align} 
For a fixed plane-wave cutoff, this also gives the relationship $\f \sim 1/N_S$ where $N_S$ is the number of subspaces in the partition.

\subsubsection{Extension to Full-Frequency Calculations} \label{fullfreq}

As noted at the end of section~\ref{chiderivations}, the convergence of $\chi^0$ was only proved for small $\omega$ roughly bounded by $\min\big(E_{\mathrm{CBM}} - E_{v=N_P^v},\allowdisplaybreaks E_{c=N_P^c} - E_{\mathrm{VBM}}\big)$. To extend this to larger $\omega$ we must treat the full Adler-Wiser formula with both its retarded and advanced components:
\begin{align}
\chi^{0, r/a}_{\G,\G'}(\vb{q}, \omega) &= \sum_{vc\vb{k}} M_{vc\vb{k}}(\vb{q},\G)^*M_{vc\vb{k}}(\vb{q},\G')\times\frac{1}{2}\left[\ddfrac{1}{E_{v\vb{k}+\vb{q}} - E_{c\vb{k}} - \omega \mp i\delta} + \ddfrac{1}{E_{v\vb{k}+\vb{q}} - E_{c\vb{k}} + \omega \pm i\delta }\right],\label{adler-wiser-full}
\end{align}
where the upper (lower) signs are for the retarded (advanced)
function, and $\delta$ is a broadening used in practice to account for finite k-point sampling. Eq.~(\ref{adler-wiser-full}) has symmetric poles at both positive and negative $\omega$, and both terms individually take the same form as the $\chi^0(\omega=0)$ in Eq.~(\ref{adler-wiser}). Thus, for finite $\omega\neq 0$, we can see that the same derivations from before go through, as long as we don't produce any divergent terms. Due to the broadening $i\delta$, we never encounter true divergences, so convergence is still achieved in a similar manner as for the Green's function in section S\ref{greenconvderivation}. 

While convergence is achieved, it is not as rapid as for the GPP due to $\omega$ being in the regime of poles which are approximated with pseudobands. To make convergence more rapid, we employ the following scheme to modify the subspace partition described in the previous section: Given a maximum frequency of interest $\omega_\max$ and a frequency-grid spacing $\delta \omega$ over which the dielectric function is to be sampled, we partition subspaces so that they span a uniform energy range $\Delta E_S = \mathrm{const.}$ up to $\omega_\max$. Beyond $\omega_\max $, we begin the exponential slices as usual. We typically take $\Delta E_S \sim \delta \omega$ for the uniform slices, though the energy range of each uniform slice is an additional convergence parameter that only appears for these full-frequency calculations. The full-frequency partitioning scheme is depicted diagrammatically in Figure \ref{fullfreq_diagram} below.

\begin{figure}[H]
    \centering
    \includegraphics[width=\textwidth]{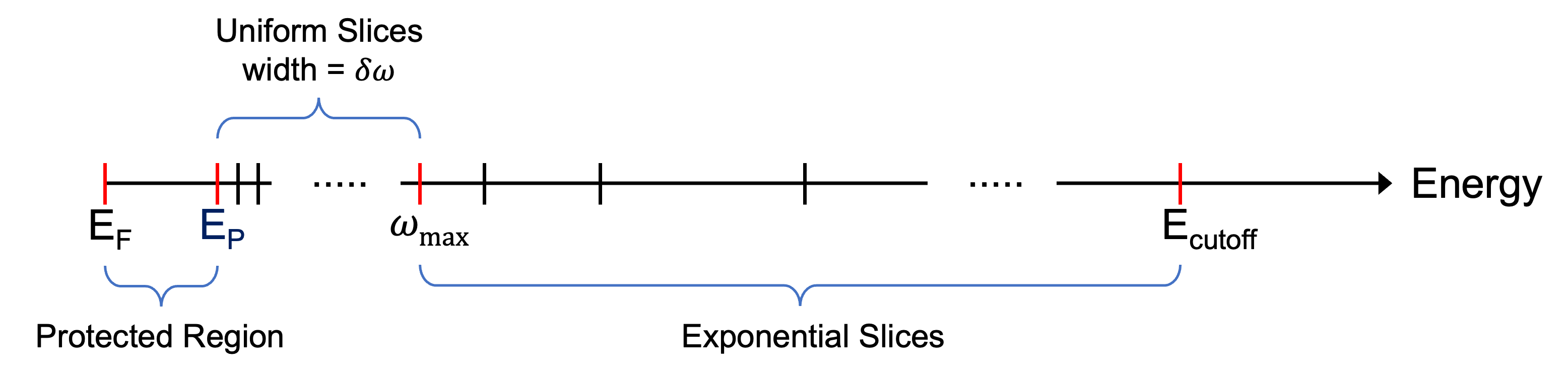}
    \caption{Diagram of the slice partition for pseudobands that can handle full-frequency dielectric calculations. Only the partition for the unoccupied states is shown, but the same is true for the occupied states.}
    \label{fullfreq_diagram}
\end{figure}

We note that $\omega = 0$ is no longer a special frequency in fully frequency-dependent calculations that do not rely on plasmon-pole models, hence the uniform slices to control the error induced from being in the vicinity of poles. We can extend this argument to deduce that, in either the fully frequency-dependent or plasmon-pole-based calculations, we may set $N_P=0$ and still achieve convergence with no protected states. As $N_S$ and $N_\xi$ increase, the error of the polarizability of the slice at $E_F$ still tends to 0. This allows a fully quadratic formalism for the evaluation of the polarizability, though the drawback is that one must use better convergence parameters than if $N_P$ was finite. In practice, we used finite values for $N_P$ to simplify the workflow, since it allows us to use a similar set of wavefunction files for the calculation of the polarizability and self-energy, though this was not required.

Moreover, this argument for setting $N_P = 0$ extends to metals. In particular, the reason for changing the way we choose slices in the explicit frequency scheme is because we evaluate the polarizability at energies $\omega$ that are close to occupied-to-unoccupied transitions that appear as poles in the polarizability. For metals, either within plasmon-pole models or an explicit frequency-dependent calculation, one runs into the same issue of evaluating the polarizability near a pole, except this pole occurs at $\omega = 0$. However, as argued above, with the use of a finite broadening $\delta$, the pseudobands approach always converges for the evaluation of any frequency. Combined with the fact that one can always use a finite (convergable to 0) broadening $\delta$ either in plasmon-pole or explicit frequency calculations, we conclude that $N_P = 0$ is applicable to metals as well as semiconductors. In fact, a broadening not smaller than $\delta \sim v_F \Delta \textbf{k}$ is desirable anyway (where $v_F$ is the Fermi velocity and $\Delta \textbf{k}$ is the $\textbf{k}$-grid spacing), since using a smaller broadening artificially discretizes the density of states and does not capture the continuum of transitions which characterizes the dielectric response of metals.

\subsection{Convergence of the GW Self-Energy $\Sigma^{GW}$}
The self-energy is somewhat more complicated to evaluate by hand, and we only discuss convergence qualitatively. However, having shown convergence for $G[\mathrm{SPB}]$ and $\chi^0[\mathrm{SPB}]$, convergence of $\Sigma^{GW}$ follows.

Here, we review the expressions used in the contour-deformation approach for computing $\Sigma^{GW}$ that is commonly employed in practice~\cite{oschlies1995gw, lebegue2003implementation, bruneval2005exchange}. We analyze the forms of these expressions and make some important points for applying pseudobands to $\Sigma^{GW}$. The expressions evaluated for $\Sigma^{GW}$ in the contour-deformation approach are
\begin{align}
    \Sigma^{GW}(\vb{r},\vb{r'},\omega) &\equiv \Sigma^{X}(\vb{r},\vb{r'}) + \Sigma^{\mathrm{Cor}}(\vb{r},\vb{r'},\omega) \\
    \Sigma^{\mathrm{Cor}}(\vb{r},\vb{r'},\omega) &\equiv \Sigma^{\mathrm{Int}}(\vb{r},\vb{r'},\omega) + \Sigma^{\mathrm{Res}}(\vb{r},\vb{r'},\omega)\\
    \Sigma^{X}(\vb{r},\vb{r'}) &= -\sum_v^{\mathrm{occ}} \phi_v(\vb{r})\phi^*_v(\vb{r}')v(\vb{r},\vb{r'})\\
    \Sigma^{\mathrm{Res}}(\vb{r},\vb{r'},\omega) &= -\sum_n \phi_n(\vb{r})\phi^*_n(\vb{r}') W^{\mathrm{Cor}}(\vb{r},\vb{r'}, E_n - \omega) \times\left[f_n\theta(E_n-\omega) - (1-f_n)\theta(\omega-E_n)\right]\\
    \Sigma^{\mathrm{Int}}(\vb{r},\vb{r'},\omega) &= -\frac{1}{\pi}\sum_n \phi_n(\vb{r})\phi^*_n(\vb{r}')\int_0^\infty d\omega'\ddfrac{\omega-E_n}{\left(\omega-E_n\right)^2 + \left(\omega '\right)^2}\times W^{\mathrm{Cor}}(\vb{r},\vb{r'}, i\omega'),
\end{align}
where $W^{\mathrm{Cor}} \equiv W(\vb{r},\vb{r'},\omega) - v(\vb{r},\vb{r'})$ , $v$ is the bare Coulomb interaction, $W = \epsilon^{-1}v$ is the screened Coulomb interaction, $f_n$ is the Fermi-Dirac occupation factor, and $\theta(\omega)$ is the Heaviside step function.

We note that $\Sigma^{\mathrm{Int}}$ involves an integral of the screened Coulomb interaction along the imaginary axis, so it does not pick up poles of $W$ or $G$~\cite{bruneval2005exchange}, and one can directly utilize the pseudobands approach. On the other hand, when evaluating the self-energy for states close to the Fermi energy, $\Sigma^{\mathrm{Res}}(\omega)$ involves sums over final states $n$ having energy $E_n$ \emph{closer} to the Fermi energy than the energy $\omega$ at which one evaluates the self-energy~\cite{bruneval2005exchange}. Such final states depend sensitively on the energy at which one is approximating them, which one can account for by systematically using smaller broadening parameters and stochastic subspaces, as in the case of metals, or by simply employing a nonzero protection window $N_P$. Because self-energy calculations are relatively inexpensive, and because we eventually wish to compute matrix elements $\bra{n\textbf{k}} \Sigma^{GW}\ket{n\textbf{k}}$ evaluated on deterministic states $\ket{n\textbf{k}}$, we find it easier to simply employ a finite $N_P$.
Finally, we note that the exchange interaction $\Sigma^{X}$ is static but involves sums over all occupied states. While we can in principle benefit from the stochastic pseudobands approach, we notice that the speed-up for the systems studied here is not advantageous given the extra stochastic error. So, the pseudobands approach is also valid for the evaluation of the quasiparticle self-energy, although we recommend one to use a finite value of $N_P$ and only conduction pseudobands. While this approach may be optimized in the future, the evaluation of the self-energy for the VBM and CBM scales as $O(N^3)$, and hence does not represent the computational bottleneck in typical GW calculations.

This concludes the convergence derivations of GW quantities within the pseudobands approach.

\bibliography{bibliography}